\documentclass[sigconf]{acmart}

\usepackage{tabu} 
\usepackage{xcolor}
\usepackage{subcaption}
\usepackage{multirow}
\usepackage{float}
\usepackage{cleveref}
\usepackage{soul}
\usepackage{xcolor,colortbl}
\usepackage{pifont}
\newcommand{\revision}[1]{\textcolor{black}{#1}}

\newcommand{\rebuttal}[1]{\textcolor{black}{#1}}

\AtBeginDocument{%
  \providecommand\BibTeX{{%
    \normalfont B\kern-0.5em{\scshape i\kern-0.25em b}\kern-0.8em\TeX}}}

\copyrightyear{2024}
\acmYear{2024}
\setcopyright{acmlicensed}\acmConference[UIST '24]{The 37th Annual ACM Symposium on User Interface Software and Technology}{October 13--16, 2024}{Pittsburgh, PA, USA}
\acmBooktitle{The 37th Annual ACM Symposium on User Interface Software and Technology (UIST '24), October 13--16, 2024, Pittsburgh, PA, USA}
\acmDOI{10.1145/3654777.3676333}
\acmISBN{979-8-4007-0628-8/24/10}

\begin{document}
\title[StreetNav: Leveraging Street Cameras to Support Precise Outdoor Navigation for Blind Pedestrians]{StreetNav: Leveraging Street Cameras to Support\\ Precise Outdoor Navigation for Blind Pedestrians}


\author{Gaurav Jain}
\affiliation{
  \institution{Columbia University}
  \city{New York}
  \state{NY}
  \country{USA}
}

\author{Basel Hindi}
\affiliation{
  \institution{Columbia University}
  \city{New York}
  \state{NY}
  \country{USA}
}


\author{Zihao Zhang}
\affiliation{
  \institution{Columbia University}
  \city{New York}
  \state{NY}
  \country{USA}
}

\author{Koushik Srinivasula}
\affiliation{
  \institution{Columbia University}
  \city{New York}
  \state{NY}
  \country{USA}
}

\author{Mingyu Xie}
\affiliation{
  \institution{Columbia University}
  \city{New York}
  \state{NY}
  \country{USA}
}

\author{Mahshid Ghasemi}
\affiliation{
  \institution{Columbia University}
  \city{New York}
  \state{NY}
  \country{USA}
}

\author{Daniel Weiner}
\authornotemark[1]
\affiliation{
  \institution{Lehman College}
  \city{New York}
  \state{NY}
  \country{USA}
}

\author{Xin Yi Therese Xu}
\authornotemark[1]
\affiliation{
  \institution{Pomona College}
  \city{Claremont}
  \state{CA}
  \country{USA}
}

\author{Sophie Ana Paris}
\authornotemark[1]
\affiliation{
  \institution{New York University}
  \city{New York}
  \state{NY}
  \country{USA}
}

\author{Michael Malcolm}
\authornote{Work done during internship at Columbia University.}
\affiliation{
  \institution{SUNY at Albany}
  \city{Albany}
  \state{NY}
  \country{USA}
}

\author{Mehmet Turkcan}
\affiliation{
  \institution{Columbia University}
  \city{New York}
  \state{NY}
  \country{USA}
}

\author{Javad Ghaderi}
\affiliation{
  \institution{Columbia University}
  \city{New York}
  \state{NY}
  \country{USA}
}

\author{Zoran Kostic}
\affiliation{
  \institution{Columbia University}
  \city{New York}
  \state{NY}
  \country{USA}
}

\author{Gil Zussman}
\affiliation{
  \institution{Columbia University}
  \city{New York}
  \state{NY}
  \country{USA}
}

\author{Brian A. Smith}
\affiliation{
  \institution{Columbia University}
  \city{New York}
  \state{NY}
  \country{USA}
}

\renewcommand{\shortauthors}{Jain et al.}

\begin{abstract}
Blind and low-vision (BLV) people rely on GPS-based systems for outdoor navigation. GPS's inaccuracy, however, causes them to veer off track, run into obstacles, and struggle to reach precise destinations. While prior work has made precise navigation possible indoors via hardware installations, enabling this outdoors remains a challenge. Interestingly, many outdoor environments are already instrumented with hardware such as street cameras. In this work, we explore the idea of repurposing \textit{existing} street cameras for outdoor navigation. Our community-driven approach considers both technical and sociotechnical concerns through engagements with various stakeholders: BLV users, residents, business owners, and Community Board leadership. The resulting system, StreetNav, processes a camera's video feed using computer vision and gives BLV pedestrians real-time navigation assistance. Our evaluations show that StreetNav guides users more precisely than GPS, but its technical performance is sensitive to environmental occlusions and distance from the camera. We discuss future implications for deploying such systems at scale.
\end{abstract}



\begin{teaserfigure}
    \centering
    \includegraphics[width=0.97\linewidth]{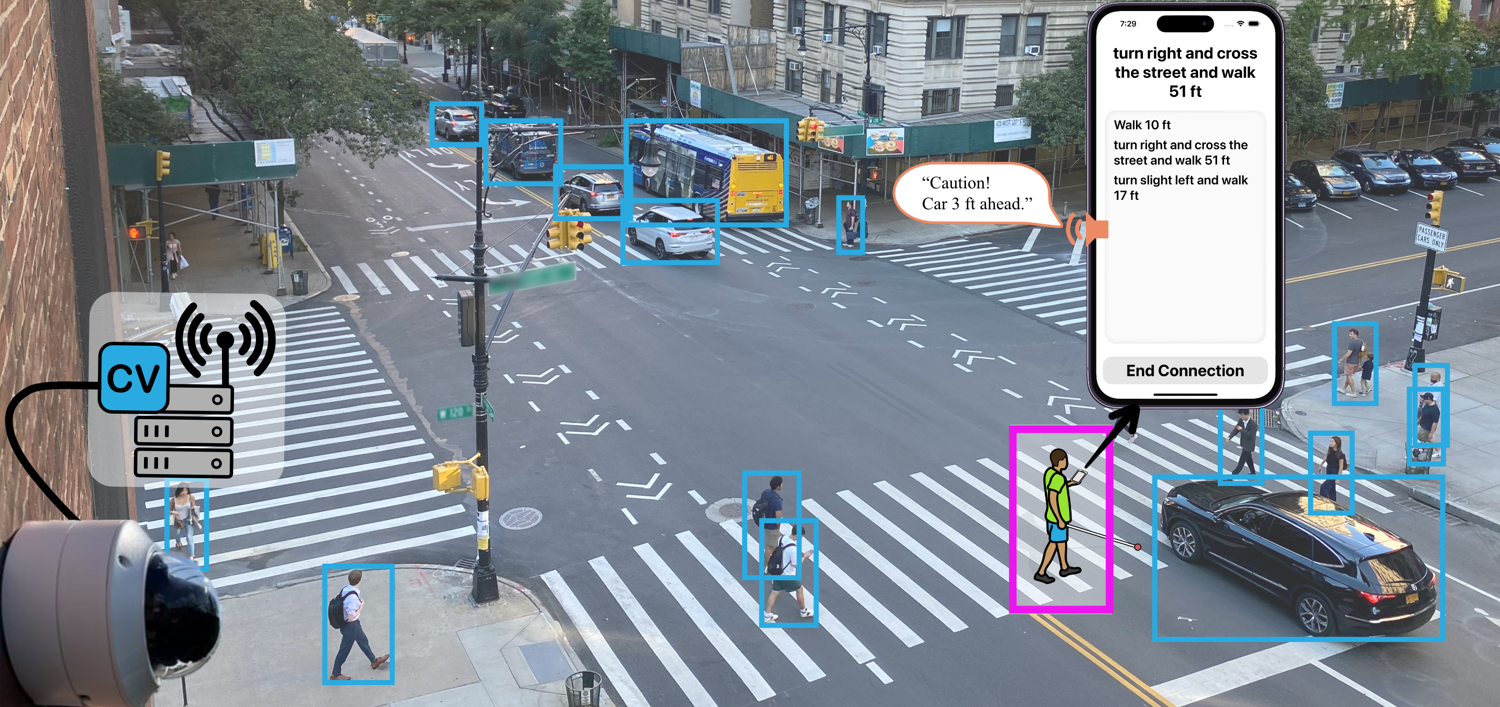}
    \caption{StreetNav is a system that explores the concept of repurposing existing street cameras to support precise outdoor navigation for blind and low-vision (BLV) pedestrians. It comprises two components: (i) a computer vision (CV) pipeline, and (ii) a companion smartphone app. The computer vision pipeline processes the street camera’s video feeds and delivers real-time navigation feedback via the app. StreetNav offers precise turn-by-turn directions to destinations while also providing real-time, scene-aware assistance to alert them of nearby obstacles and facilitate safe street crossings.
}
    \Description{Street intersection as seen from a second floor camera view. A blind pedestrian is detected by the system and is crossing the street. A car is blocking the blind pedestrian’s path, and a mobile app provides them with a warning saying “Caution! Car 3 ft ahead.”}
  \label{fig:teaser}
\end{teaserfigure}


\maketitle

\section{Introduction}
\label{sec:intro}
Outdoor navigation in unfamiliar environments is a major challenge for blind and low-vision (BLV) people. Among the many navigation systems that have been developed to assist BLV people outdoors, GPS-based systems are the most popular~\cite{blindsquare, MSsoundscape, seeingeyegps, autour, kacorri_insights_2018}. These systems, such as BlindSquare~\cite{blindsquare} and Microsoft Soundscape~\cite{MSsoundscape}, guide users to a destination and notify them of surrounding points of interest (POIs). Despite GPS's undeniable impact in making outdoor environments navigable, its imprecision is a major limitation~\cite{saha_closing_2019}. GPS precision can range from 5 meters at best to over tens of meters in urban areas with buildings and trees~\cite{gps_accuracy, modsching2006field, van_diggelen_worlds_2015}. This imprecision causes BLV people to veer off track~\cite{pan_walking_2013}, run into unexpected obstacles~\cite{pariti_intelligent_2020, presti_watchout_2019, avila_survey_2017}, and struggle to reach precise destinations~\cite{saha_closing_2019} when navigating outdoors.

Prior work on indoor navigation, on the contrary, has made precise navigation assistance possible for BLV people~\cite{ahmetovic_navcog_2016, nakajima2012light, gallagher2012sensor, kim_navigating_2016, sato_navcog3_2019}. \revision{Most approaches do so by installing a dense network of Bluetooth~\cite{ahmetovic_navcog_2016} or WiFi~\cite{gallagher2012sensor} beacons. However, extending this approach for outdoor navigation is not feasible due to the vast scale and complex nature of outdoor spaces. 
Interestingly,} many outdoor environments of interest, such as urban districts and downtown areas, are already instrumented with hardware that has the potential to help, including street cameras, traffic sensors, and other urban infrastructure components.

Street cameras, in particular, have the potential to support BLV pedestrians' outdoor navigation. The video feed from these cameras could be processed using computer vision to track BLV pedestrians and perceive their visual environment with greater precision and fidelity compared to GPS-based systems. 
The profound potential of street cameras for assistive technology is accompanied by significant challenges and concerns --- both technical and sociotechnical.

On the technical front, there is a lack of understanding regarding the precise capabilities of street cameras to track BLV pedestrians and how camera-based systems should be designed to effectively support BLV people's outdoor navigation. Sociotechnically, a major concern revolves around privacy due to cameras' capability to collect pervasive data, not only affecting BLV users but also other pedestrians and vehicles in the vicinity~\cite{emami-naeini_understanding_2023}. Moreover, street cameras are often deployed by governments to force surveillance~\cite{nypd_public_safety, ubiquitous_street_cameras, china_cams, china_us_cams, billion_cams}, which exacerbates people's privacy concerns. Limited work has been done to explore how camera-based technologies can respect people's privacy concerns and directly serve their interests, rather than solely serving government-defined purposes~\cite{emami-naeini_understanding_2023, haug_ms_thesis, drew_smart_2013, zhang_research_2019}.

In this work, we take a community-driven approach to explore the concept of leveraging street cameras to support outdoor navigation for blind pedestrians.To this end, we engage with various stakeholders including BLV users, local residents, local business owners, and Community Board leadership. We aim to address both the technical and sociotechnical aspects of this concept through the following research questions:\begin{itemize}
    \item[\textbf{RQ1.}] \rebuttal{What are stakeholders' privacy concerns toward camera-based assistive technology, and how might they be respected?}
    \item[\textbf{RQ2.}] \rebuttal{How might a street camera-based navigation assistance system be designed?}
    \item[\textbf{RQ3.}] \rebuttal{To what extent do street camera-based systems support BLV people's outdoor navigation?}
\end{itemize}


To answer RQ1, we interviewed various stakeholders, including two BLV users, two local residents, a local business owner, and a Community Board leader. We discovered stakeholders' differing perspectives on privacy concerns towards camera-based assistive technology. All stakeholders expressed that repurposing \textit{existing} cameras to help BLV people, rather than installing \textit{new} cameras, significantly alleviates their privacy concerns. Participants also shared that regulating data storage, anonymization, and access policies could further enhance their sense of comfort around privacy.

To answer RQ2, we developed \textit{StreetNav}, a system that leverages a street camera to support precise outdoor navigation for BLV pedestrians. \revision{StreetNav's design is informed by BLV people's outdoor navigation challenges (Section~\ref{sec:formative-interviews}) and by various stakeholders' privacy concerns toward camera-based assistive technology (Section~\ref{sec:formative-interviews-with-stakeholders}).}
As Figure~\ref{fig:teaser} illustrates, StreetNav comprises two components: (i) \textit{a computer vision pipeline}, and (ii) \textit{a companion smartphone app}. The computer vision pipeline processes the street camera's video feed and delivers real-time navigation assistance to BLV pedestrians via the smartphone app. StreetNav offers precise turn-by-turn directions to destinations while also providing real-time, scene-aware assistance to prevent users from veering off course, alert them of nearby obstacles, and facilitate safe street crossings. 

\rebuttal{StreetNav supports BLV pedestrians' outdoor navigation by repurposing an \textit{existing} street camera. Through StreetNav, we explore the feasibility of street camera-based systems at a single street intersection as a first step.}
We chose to use a camera from the NSF PAWR COSMOS testbed~\cite{raychaudhuri2020challenge,yang2020cosmos} because it is available to researchers after an approval process and IRB review.
We considered other publicly available testbeds, such as Mobintel~\cite{mobintel} and DataCity SMTG~\cite{datacity_smtg}, but chose COSMOS due to its location in a major city (New York) with high pedestrian and vehicle traffic. Anonymized video samples from the COSMOS cameras, including the one used in this work, can be found online~\cite{cosmos-cameras}.
To answer RQ3, we conducted \revision{both a user evaluation and a technical evaluation of StreetNav.} 
Our user evaluation involved eight BLV pedestrians who navigated routes with both StreetNav and BlindSquare~\cite{blindsquare}, a popular GPS-based navigation app specially designed for BLV people. 
Our findings reveal that StreetNav offers significantly greater precision in guiding pedestrians compared to BlindSquare. Specifically, StreetNav guided participants to within an average of $2.9$ times closer to their destination and reduced veering off course by over 53\% when compared to BlindSquare. This substantial improvement was reflected in a forced ranking, where all participants unanimously preferred StreetNav over BlindSquare.

\revision{Despite an improved user experience, StreetNav's technical evaluation exposes certain limitations. We found that although StreetNav tracks pedestrians with an 82\% precision and 65\% recall at 0.5 IOU threshold, the accuracy drops significantly as the pedestrian's distance from the camera increases. The false negative rates goes up from 1\% at a distance of 5 meters to 74\% at a distance of 40 meters from the camera. Additionally, StreetNav's performance is sensitive to occlusions and distance from camera. We discuss future implications of our findings in the context of deploying street camera-based navigation systems at scale.
}

In summary, we contribute \rebuttal{(1) a study of various stakeholders' privacy concerns toward camera-based assistive technology, (2) the StreetNav system through which we explore the concept of repurposing \textit{existing} street cameras for precise outdoor navigation assistance, and (3) both a user and technical evaluation of StreetNav.}

\section{Related Work}

Our work builds on the following three main research threads:
\textit{(i)} outdoor navigation approaches, \textit{(ii)} overhead camera-based robot navigation, and \textit{(iii)} indoor navigation approaches.


\subsubsection*{\textbf{Outdoor Navigation Approaches}}

Existing approaches for outdoor navigation primarily rely on GPS-based navigation systems for guiding users to the destination and providing information about nearby POIs~\cite{blindsquare, MSsoundscape, seeingeyegps, autour, kacorri_insights_2018}. 
BlindSquare\cite{blindsquare}, for instance, utilizes the smartphone's GPS signal to determine the user's location and then provides the direction and distance to the destination, gathered from Foursquare and Open Street Map.
The GPS signal, however, offers poor precision with localization errors as big as tens of meters~\cite{modsching2006field, yoon_leveraging_2019, ahmetovic_navcog_2016, gps_accuracy}. The accuracy is lower in densely populated cities~\cite{vicek1993gps}, which is even more concerning given that a disproportionately high percentage of BLV people live in cities ~\cite{harkey2007accessible}. Despite GPS-based systems' undeniable impact on helping BLV people in outdoor navigation, their low precision and inability to provide real-time support for avoiding obstacles and veering off the path limits their usability as a standalone navigation solution. Our work attempts to investigate street cameras' potential as an alternative solution for providing precise and real-time navigation assistance.


Another approach for outdoor navigation has explored developing personalized, purpose-built, assistive devices that support BLV people with scene-aware aspects of outdoor navigation, such as crossing streets~\cite{son_crosswalk_2020, arai_cross-safe_2020, guy_crossingguard_2012}, recording routes~\cite{yoon_leveraging_2019}, and avoiding obstacles~\cite{wang_enabling_2017, duh2020v, lin_deep_2019, ran2004drishti, katzschmann2018safe, fiannaca_headlock_2014}. While these solutions address some of the precise and real-time aspects of BLV people's outdoor navigation, support for point-to-point navigation is missing. Consequently, they do not offer a comprehensive, all-in-one solution for outdoor navigation. Furthermore, these systems place the burden of purchasing devices onto the BLV users. Our work, by contrast, explores the possibility of using existing street cameras to provide a comprehensive solution for outdoor navigation. We investigate repurposing existing hardware in outdoor environments to support accessibility applications, thus directly imbuing accessibility within the city infrastructure at no additional cost to the BLV user.


\subsubsection*{\textbf{Overhead Camera-based Robot Navigation}}
A parallel research space to street cameras for blind navigation is robot navigation using overhead cameras. One common subspace within this field is sensor fusion for improved mapping. Research in this space focuses on fusing information between sighted ``guide'' robots and overhead cameras~\cite{chang_mobile_2013}, fusing multiple camera views for improved tracking~\cite{chang_mobile_2013, oscadal_smart_2020, pflugfelder_localization_2010}, and improving homography for robust mapping, independent of camera viewing angle~\cite{shim_mobile_2016, shim_mobile_2015}. 
Another challenge tackled within this space is robot path planning. Research in this space aims to improve path planning algorithms~\cite{chang_mobile_2013, oscadal_smart_2020, shim_mobile_2015}, assign navigational tasks to robot assistants~\cite{chang_mobile_2013, oscadal_smart_2020}, and address the balance between obstacle avoidance and path following~\cite{chang_mobile_2013, shim_mobile_2015}. While prior work on robot navigation using fixed cameras explores the research space of automating ``blind'' robot navigation, our work explores how fixed cameras, specifically street cameras, could be repurposed to support navigation for blind pedestrians. 
\revision{
Our preliminary work~\cite{jain_towards_streetcamera_2023} explores an initial system concept that considers street cameras for blind navigation. This concept was not evaluated, however, nor were community issues considered. In this work, we perform both a technical and user evaluation to holistically explore the concept of leveraging street cameras for blind navigation. Moreover, we take a community-driven approach to consider both technical and sociotechnical challenges in developing street camera-based navigation systems, engaging with not only BLV users but also various stakeholders.
} 

\subsubsection*{\textbf{Indoor Navigation Approaches}}
Prior work in indoor navigation assistance has made significant progress through the utilization of various localization technologies, which usually relies on hardware like WiFi or Bluetooth beacons \cite{ahmetovic_navcog_2016, nakajima2012light, gallagher2012sensor, kim_navigating_2016, sato_navcog3_2019}. These solutions have proven highly effective within indoor environments. NavCog3 \cite{ahmetovic_navcog_2016}, for example, excels in indoor navigation by employing Bluetooth beacons for precise turn-by-turn guidance. Nakajima and Haruyama \cite{nakajima2012light} exploit the use of visible lights communication technology, utilizing LED lights and a geomagnetic correction method to localize BLV users. However, extending these approaches to support outdoor navigation is not feasible. This is particularly evident when considering the substantial effort in hardware setup that these systems typically require, making them ill-suited for the larger, unstructured outdoor environment. Furthermore, most of these methods lack the capability to assist with obstacle avoidance and to prevent users from veering off course --- both of which are less severe issues indoors compared to outdoors~\cite{pan_walking_2013}. Our exploration of using existing street cameras is better suited to address the largely unaddressed challenges of outdoor navigation. This approach has the potential to offer precise localization without requiring dense hardware installations. It can harness existing street cameras for locating a pedestrian's position. Additionally, it holds the potential to tackle the distinctive challenges posed by the unstructured nature of outdoor environments, including real-time obstacle avoidance and safe street crossing.


\section{BLV People's Challenges in Outdoor Navigation Using GPS-based Systems}
\label{sec:formative-interviews}
\rebuttal{We conducted semi-structured interviews with six BLV participants to identify challenges that they face when navigating outdoors using GPS-based systems. Our interviews found three major challenges, \textbf{C1}: following routing instructions through complex environment layouts, \textbf{C2}: avoiding unexpected obstacles while using GPS-based systems, and \textbf{C3}: crossing streets safely. While these challenges are well-documented within existing literature~\cite{pan_walking_2013, pariti_intelligent_2020, presti_watchout_2019, avila_survey_2017, saha_closing_2019}, our findings highlight areas that could be prioritized for resolution through the implementation of a street camera-based navigation system. 
Appendix~\ref{sec:formative_study_appendix} provides additional detail on participant demographics, interview procedure, and interview findings.}

\section{Stakeholders' Privacy Concerns Toward Camera-Based Systems}

\label{sec:formative-interviews-with-stakeholders}

\revision{
We conducted five semi-structured interviews with various stakeholders from Harlem, New York City, where the COSMOS testbed is located. Harlem is a diverse community within a major city that has become sensitive to government surveillance and overreach. The interviews were with two BLV users (B1, B2), two local residents (R1, R2), a local business owner (O1), and a Community Board leader (CB1). 
Our objective was to understand stakeholders' privacy concerns regarding camera-based assistive technology and explore ways to address these concerns (RQ1).
}

\subsection{Methods}

\subsubsection*{\textbf{Participants}}
\revision{
\rebuttal{Table~\ref{tab:stakeholder-study-ptcpts} (Appendix~\ref{sec:ptcpt-demographics}) reports participant demographics.} Each interview lasted for about 45-60 minutes, except for the interview with the Community Board leader that lasted for 15 minutes. Three interviews (B1, B2, R1) were conducted online over Zoom, two (O1, R1) were conducted in person, and one (CB1) was conducted over phone. All participants, except for CB1, were compensated \$50 for their participation in this IRB-approved study. CB1 refused to accept the compensation.
}



\subsubsection*{\textbf{Procedure}}
\rebuttal{
We began by giving participants a short presentation describing an initial system concept. The presentation illustrated how street cameras could capture street intersections, use computer vision to track pedestrians and vehicles, and deliver navigation instructions to BLV users via smartphones. 
We verbally described visuals to BLV participants during the study.
We then asked participants questions about their perceived benefits and concerns, preferences around data collection and use scenarios that may raise privacy concerns: e.g., \textit{Does it matter to you who has access to the camera feed?} 
During interview with the Community Board leader, we inquired about the feasibility of such a system: e.g., \textit{How feasible would it be to use street cameras for assistive technology purposes?} We concluded interviews by discussing strategies for how such systems might respect their privacy concerns.
}

\subsubsection*{\textbf{Interview Analysis}}
\revision{
We used thematic analysis~\cite{braun_using_2006} to analyze the interviews, similar to our methodology described in Section~\ref{sec:formative-interviews}. This analysis involved three researchers independently generating initial sets of codes, which were then collaboratively iterated to identify emerging themes.
}

\subsection{Findings: Privacy Concerns}

\revision{Our participants, irrespective of their stakeholder category, held differing perspectives on privacy concerns toward camera-based assistive technology. While some had no privacy concerns whatsoever, others felt uncomfortable with the concept of a camera monitoring them. When asked if there was anything that could satisfy their concerns, concerned participants identified two strategies: \textit{(i)} regulating data storage, anonymization, and access policies; and \textit{(ii)} repurposing  \textit{existing} cameras rather than installing \textit{new} cameras to assist BLV people. The following sections detail our findings on stakeholders' differing viewpoints on privacy and strategies that this assistive technology could employ to respect those viewpoints.}

\subsubsection*{\textbf{Stakeholders' differing perspectives on privacy concerns.}} 
%

\revision{Nearly half of the participants (B1, R2, O1) expressed no concerns about being recorded by the camera. In fact, they highlighted the added benefits of street cameras in enhancing public safety, particularly aiding in crime investigation. 
These participants expressed the willingness to sacrifice some privacy in exchange for societal benefits such as accessibility and public safety. This finding aligns with earlier findings by Profita et al.~\cite{profita_at_2016}. Additionally, B1 pointed out that complete privacy should not be expected in public spaces: ``\textit{If you're on a public street, you pretty much could expect for anyone to see you at any time. So it's no more invasive than anything else on a public street.
A public street is pretty much fair game for anybody.}'' 
}

\revision{
In contrast, other participants (B2, R1) expressed discomfort with cameras' capability not only to track people's movements but also to ``\textit{know what [they] look like}'' (B2). R1 compared a camera's presence to an ``\textit{overarching shadow that's always looking over [and] monitoring their everyday moves.}'' 
These participants voiced concerns against the use of such cameras for public safety purposes. They feared that the ability to determine individuals' identities from the video feed could result in the targeting of marginalized groups such as people of color (R1) and BLV individuals (B2). As B2 stated: ``\textit{The fact that I'm being surveilled even more as a blind person, and knowing that police disproportionately target the disabled whenever things are going wrong, that just makes me feel even less safe.}''
}

\subsubsection*{\textbf{Regulating data storage, anonymization, and access policies.}}
\revision{
We inquired about participants' preferences regarding the collection and storage of the video feed. Those without privacy concerns (B1, R2, O1) expressed indifference regarding the duration and form (e.g., anonymized vs. raw footage) of video footage storage, asserting they had ``\textit{nothing to hide}'' (O1). B1 elaborated: ``\textit{It really doesn't matter to me. I'm not the person who is going to commit the crime, so I don't care if they keep [the data] forever.}'' 
Conversely, participants with privacy concerns (B2, R1) expressed discomfort with any long-term data storage if such storage was not necessary for the functionality of the assistive technology. They proposed anonymizing the video footage by techniques such as blurring faces (R1) or representing individuals with ``dots'' (B2) or avatars akin to those used in GPS-based applications. However, complete anonymization would negate safety benefits desired by some individuals. A compromise was reached in favor of limited storage duration, up to a week, alongside clear guidelines regarding access and legal use of the video footage. Most participants expressed greater trust in government entities than in corporations to manage these cameras, citing concerns about potential data exploitation by the latter.
}

\subsubsection*{\textbf{Repurposing existing cameras rather than install new cameras}}
\revision{
During the interview, R2 highlighted the ubiquity of cameras in urban areas: ``\textit{It's New York, there's going to be a camera every other block. There's no way that these cameras can't pick you up.}'' We pursued this observation with other participants and discovered that assisting BLV pedestrians with \textit{existing} cameras rather than install \textit{new} cameras significantly alleviated their privacy concerns.
B2 affirmed this, saying, ``\textit{I would be okay with that because, you know, it's a dual purpose thing. [The Dept. of Transportation] is already putting the speeding cameras there, so at least it does something nice for people while the camera is in place.}'' 
The business owner, O1, consented to lending the cameras at their restaurant's entrance, overlooking the street, under two conditions: \textit{(i)} it should be used solely and responsibly to assist people, and \textit{(ii)} it should not record any views inside their restaurant.
} 


\revision{
We consulted with the Community Board leader (CB1) to understand the feasibility of repurposing existing street cameras. CB1 emphasized the need for collaboration among various government entities to effectively enable this technology. This collaboration would not only involve granting access to the cameras but also ensuring that they possess the necessary capabilities to support this application. CB1 identified several key institutions that could play vital roles in this effort: the Department of Transportation, responsible for providing camera access and relevant technical support; the Department of Buildings or the Metropolitan Transportation Authority (MTA), tasked with granting camera access and permissions to house any required computational resources; and the National Security Agency (NSA), tasked with ensuring that camera access maintains security protocols. Additionally, CB1 highlighted the importance of implementing processes to monitor the impact of this technology on local communities. For instance, public outreach initiatives would help the public understand the purpose of the technology, ensuring transparency and accountability throughout the deployment process.
}

\section{The StreetNav System}
\revision{StreetNav is a system that explores the concept of repurposing \textit{existing} street cameras to support outdoor navigation for BLV pedestrians (RQ2). The following sections describe StreetNav's design rationale (Section~\ref{sec:design-rationale}), the computer vision pipeline (Section~\ref{sec:cv-pipeline}), and the smartphone app's user interface (Section~\ref{sec:user-interface}).} 

\subsection{StreetNav: Design Rationale}
\label{sec:design-rationale}
\revision{Our design and development of StreetNav considers prior work on navigation assistance, functions of traditional mobility aids, and insights gathered from our interviews with BLV people (Section~\ref{sec:formative-interviews}) and with various stakeholders (Section~\ref{sec:formative-interviews-with-stakeholders})}

\revision{
To address challenges that BLV people face when navigating outdoors using existing GPS-based systems, StreetNav provides users precise turn-by-turn navigation instructions to destinations and prevents veering off track (\textbf{C1}); gain awareness of nearby obstacles (\textbf{C2}); and assist in crossing streets safely (\textbf{C3}). StreetNav enables these affordances through its two main components: (i) \textit{computer vision pipeline}, and (ii) \textit{companion smartphone app}. The computer vision pipeline processes the street camera’s video feeds to give BLV pedestrians real-time navigation feedback via the app.
}

\begin{figure}[t]
    \centering
    \includegraphics[width=0.9\linewidth]{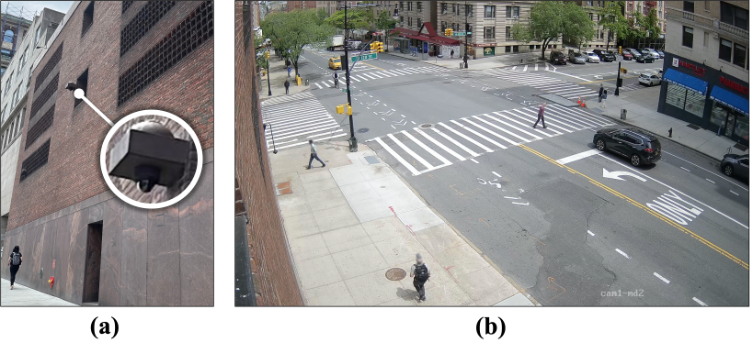}
    \caption{Street camera used for StreetNav's development and evaluation. The camera is (a) mounted on the building's second floor and (b) faces a four-way intersection.}
    \Description{Two part image labeled (a), and (b). Described from left to right: (a) left side shows a picture of the camera used for the system and mounting position on the side of the building. (b) right side shows a four way intersection from the point of view of the camera.
}
    \label{fig:camera_location_view}
\end{figure}

\revision{
To ensure that StreetNav respects stakeholders' privacy concerns, we explore how an \textit{existing} camera may be repurposed, rather than installing a \textit{new} camera, to support BLV people's outdoor navigation (Section~\ref{sec:formative-interviews-with-stakeholders}). For this reason, we chose a camera that faces a four-way street intersection---the most common type of intersection---and is mounted on a building's second floor, offering a typical street-level view of the intersection. Figure~\ref{fig:camera_location_view} shows the street camera and its view of the street intersection. 
StreetNav eliminates the requirement of \textit{storing} any video data by processing the camera feed in real-time to generate navigation instructions.}

\rebuttal{Appendix~\ref{sec:technical-setup} describes StreetNav's technical setup which enables the real-time navigation assistance.}

\begin{figure}[t]
    \centering
    \includegraphics[width=0.95\linewidth]{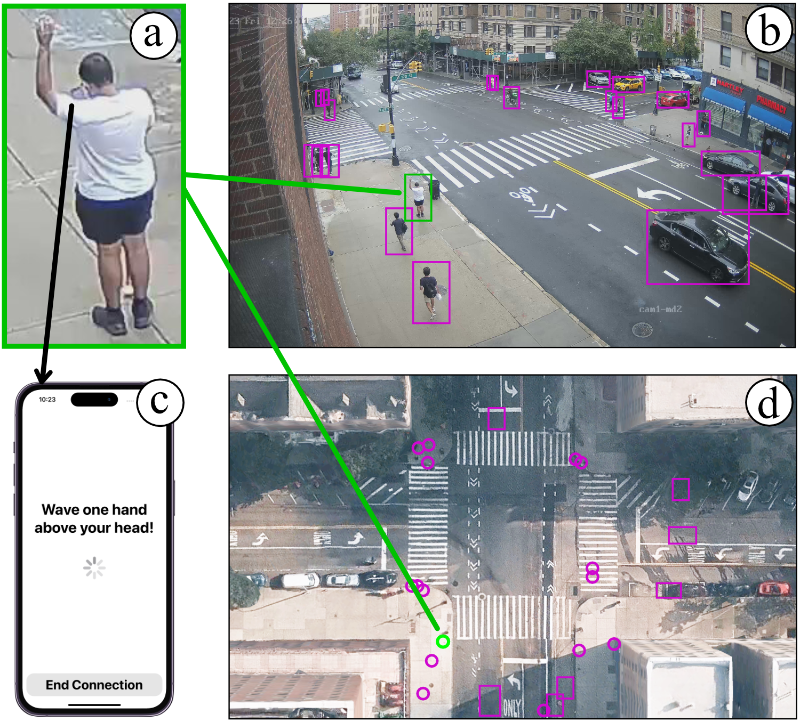}
         \caption{Gesture-based localization for determining a user's position on the map. (a) A study participant (P1) is (c) prompted to wave one hand above their head, enabling the computer vision pipeline to distinguish them from other pedestrians in (b) the camera feed view and (d) the map. 
         }
         \Description{Four part image labeled (a), (b), (c), (d). (a) Top left is a blind man standing on the sidewalk, waving one hand above his head, with a phone in the other hand. (c) Bottom left is a screenshot of a phone screen prompting the user to wave. (b) Top right is a street intersection with bounding boxes around all of the pedestrians in the image. (d) Bottom right is the map view of the same intersection with circles indicating the positions of pedestrians.}
         \label{fig:localization}
\end{figure}

\subsection{StreetNav: Computer Vision Pipeline}
\label{sec:cv-pipeline}
StreetNav's computer vision pipeline processes the street camera's video feed in real time to facilitate navigation assistance. It consists of four components: (i) \textit{localizing and tracking the user}: locating user's position on the environment's map; (ii)~\textit{planning routes}: generating turn-by-turn navigation instructions from user's current position to destinations; (iii) \textit{identifying obstacles}: predicting potential collisions with other pedestrians, vehicles, and objects (e.g., trash can, pole); and (iv) \textit{recognizing pedestrian signals}: determining when it is safe for pedestrians to cross (walk vs. wait) and calculating the duration of each cycle. Next, we describe the computer vision pipeline's four components in detail.

\subsubsection*{\textbf{Localizing and tracking the user}}
To offer precise navigation assistance, a system must first determine the user's position from the camera view and then project it onto the environment's map. Figure~\ref{fig:localization}d shows the map representation we used, which is a snapshot from Apple Maps'~\cite{applemaps} satellite view of the intersection where the camera is deployed.

StreetNav tracks pedestrians from the camera's video feed using Nvidia's DCF-based multi-object tracker~\cite{nvidia_dcf_tracker} and the YOLOv8 object detector~\cite{terven2023comprehensive}. 
The tracker detects all pedestrians and assigns them a unique ID. However, the system needs a way to differentiate between the BLV user and other pedestrians.

Figure~\ref{fig:localization} shows the \textit{gesture-based localization} approach we introduced to address this issue. To connect with the system, BLV pedestrians must wave one hand above their head for 2--3 seconds (Figure~\ref{fig:localization}a), enabling the system to determine the BLV pedestrian's unique tracker ID. We chose this gesture after discussions with several BLV individuals, including our BLV co-author, and most agreed that this single-handed action was both convenient and socially acceptable to them. Moreover, over-the-head gestures such as waving a hand can also be detected when users are not directly facing the street camera. 

We implement hand gesture-based localization by first creating image crops of all detected pedestrians, then classifying them as `waving' or `walking' pedestrians using CLIP~\cite{radford_learning_2021}. CLIP classifies each pedestrian by computing visual similarity between the pedestrian's image crop and two language prompts: `person walking' and `person waving hand.' We experimentally fine-tuned the confidence thresholds and these language prompts. 

\revision{We estimate the user's feet position to be the mid-point of bounding box's bottom edge. Finally, we transform the user's feet position from the street camera view (Figure~\ref{fig:localization}b) to the map (Figure~\ref{fig:localization}d)} using a simple feed-forward neural network trained on data that we manually annotated. The network takes as input the 2D pixel coordinate from the street camera view and outputs the corresponding 2D coordinate on the map. 

\subsubsection*{\textbf{Planning routes}}
\begin{figure}[t]
    \centering
    \includegraphics[width=0.95\linewidth]{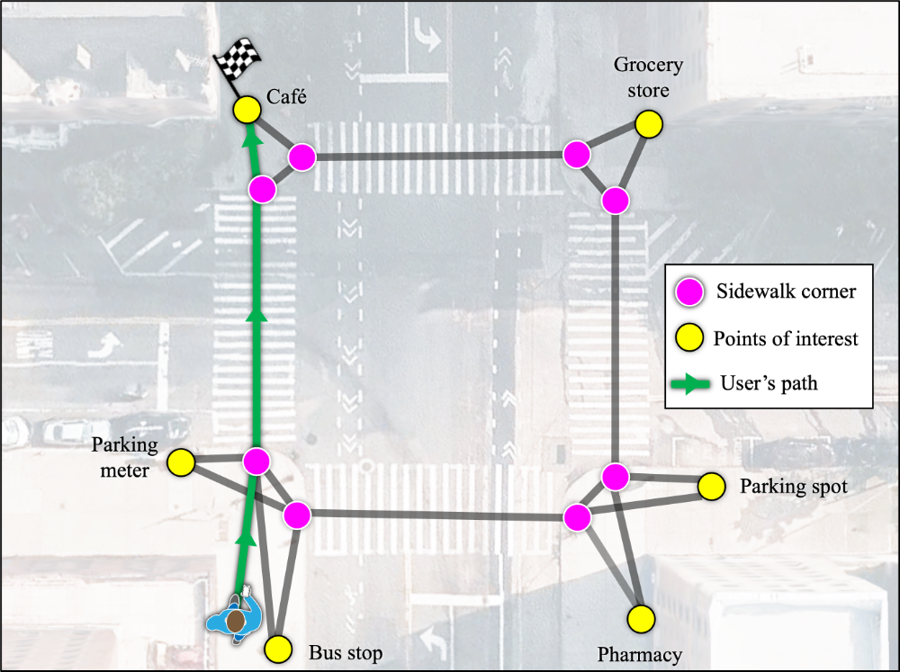}
         \caption{StreetNav's internal graph representation for route planning. The user's current position is added dynamically as a start node to the graph upon choosing a destination. The shortest path, highlighted in green, is then calculated as per this graph representation. 
         }
         \Description{A birds eye view of the intersection with grey lines representing the internal graph for route planning and a green line indicating the current route of the user. There are pink and yellow dots along the green graph lines, the pink dots indicating sidewalk corners and yellow dots indicate points of interest.}
         \label{fig:graph}
\end{figure}

\rebuttal{To plan routes, a street camera-based systems require a map of the environment, internally represented as a graph with waypoints and connections between them. For StreetNav, one of the researchers manually annotated a satellite view image of the street intersection to create this graph, a process that took roughly 10 minutes. This process could be automated by integrating with OpenStreetMap~\cite{osm} map data in the future.}

Figure~\ref{fig:graph} shows the internal graph structure that StreetNav uses for planning routes. Similar representations have been used in prior work on indoor navigation systems~\cite{ahmetovic_navcog_2016, sato_navcog3_2019, guerreiro_cabot_2019}. In the graph, nodes correspond to POIs and sidewalk corners, whereas edges correspond to walkable paths. Once the user chooses a destination from the POIs, StreetNav adds the user's current position as a start node to this graph representation and computes the shortest path to the chosen POI using A$^{*}$ algorithm~\cite{duchon_path_2014}. 

\subsubsection*{\textbf{Identifying obstacles}} 

StreetNav provides users with information about an obstacle's category and relative location. This gives users context on the size, shape, and location of an obstacle; enabling them to confidently apply their mobility skills to go around unexpected obstacles.

\begin{figure}[t]
    \centering
    \includegraphics[width=0.99\linewidth]{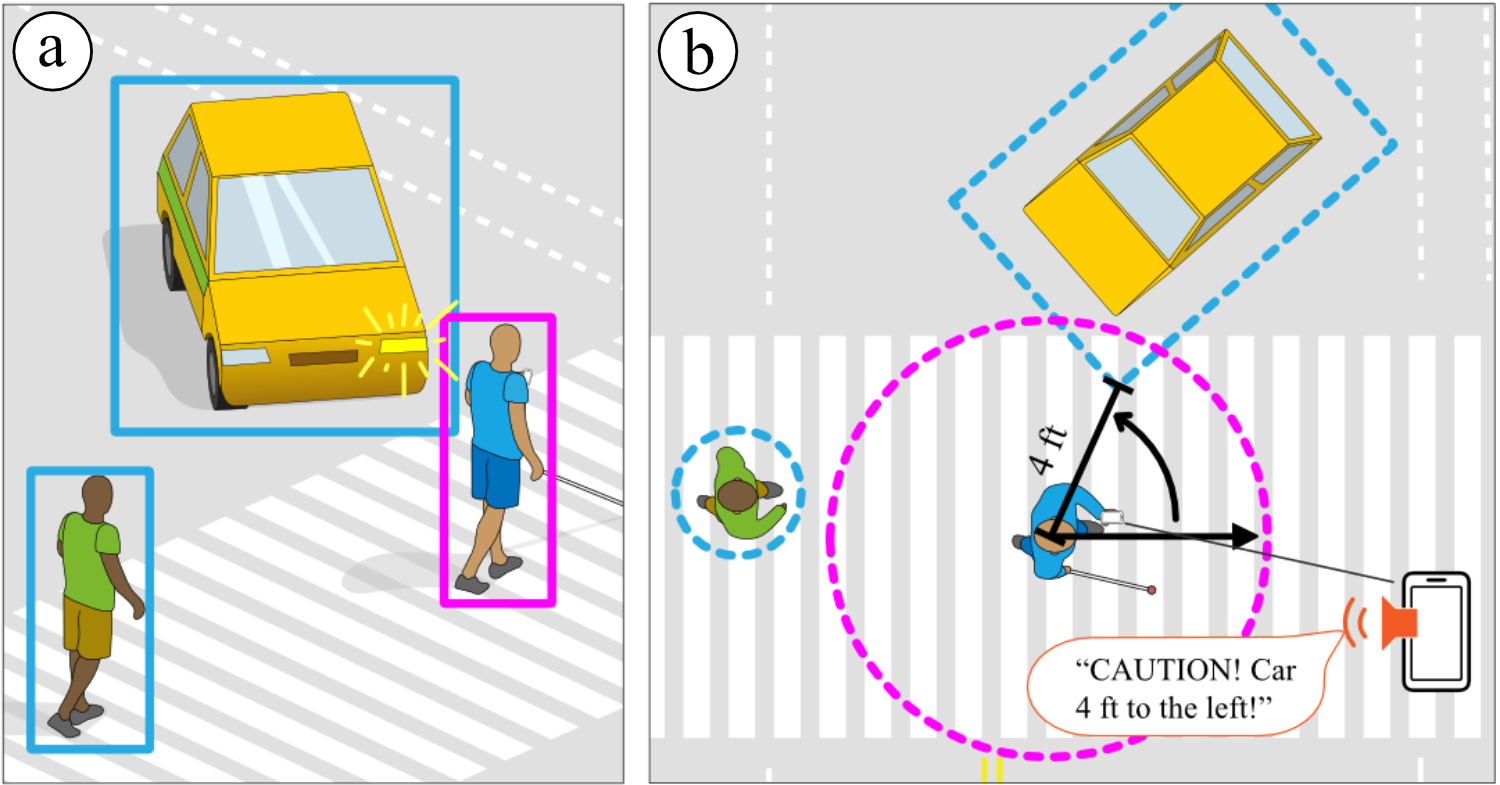}
         \caption{Identifying obstacles in the user's vicinity. (a) A vehicle turning left yields to the BLV pedestrian (detected in purple) crossing the street. (b) StreetNav identifies the obstacles' category and relative location on the map to provide real-time feedback via the app.}
         \Description{Side by side, two image views of a BLV pedestrian crossing the street with a car turning towards the pedestrian. The left image is from a simulated street camera’s POV, the right image is a bird’s eye view with circles around the user and the car, indicating the car is 4 feet away from the pedestrian. The pedestrian’s phone in the corner of the image is shown, alerting the pedestrian of the oncoming car.}
         \label{fig:obstacle-awareness}
\end{figure}

Figure~\ref{fig:obstacle-awareness} illustrates how the system identifies obstacles in the user's vicinity. StreetNav's multi-object tracker is used to track other objects and pedestrians. Examples of other objects include cars, bicycles, poles, and trash cans. The computer vision pipeline then projects the detected objects' positions onto the map. To identify obstacles in the BLV user's vicinity, StreetNav computes the distance and angle between the user and other detected objects with respect to the map (Figure~\ref{fig:obstacle-awareness}b). Any object (or pedestrian) within a fixed radial distance from the BLV user is flagged as an obstacle. Through a series of experiments with our BLV co-author, we found that a $4$ foot radius works best for StreetNav to provide users with awareness of obstacles in a timely manner.

\begin{figure}[t]
    \centering
    \includegraphics[width=0.99\linewidth]{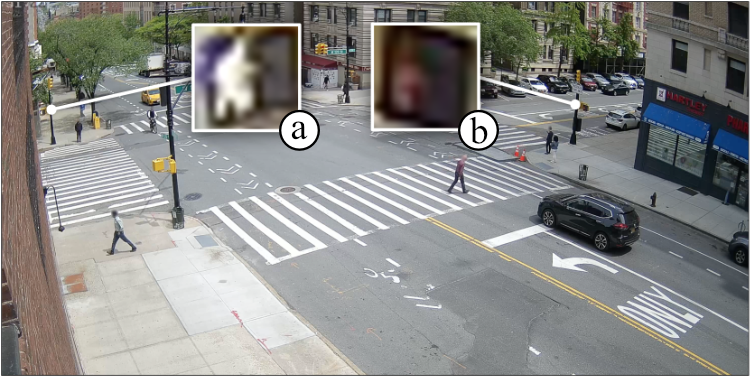}
         \caption{Recognizing pedestrian signal states. StreetNav compares the number of white and red pixels in the signal crops to determine its state: (a) \textit{walk} vs. (b) \textit{wait}.}
         \Description{Street intersection with two zoomed in frames showing the crosswalk signals from the street camera’s point of view, (a) is of a walk signal, (b) is of a wait signal.}
         \label{fig:pedestrian-signals}
\end{figure}

\subsubsection*{\textbf{Recognizing pedestrian signals}}
To determine the pedestrian signals' state (i.e., \textit{walk} vs. \textit{wait}), we leverage the fact that walk signals are always white and wait signals are always red in color. \rebuttal{A street camera-based system would first need to detect pedestrian signals from the camera feed before detecting its state. For StreetNav's implementation, one of the researchers manually annotated the pedestrian signals' screens in the camera feed. Future iterations could scale this process by automatically detecting signals by training custom object detectors.} 

Figure~\ref{fig:pedestrian-signals} shows pedestrian signals in the camera's video feed. StreetNav applies pixel-thresholding onto the pedestrian signal crops to filter all white and red pixels. Then, it compares the number of white and red pixels to determine signal state: \textit{walk} (Figure~\ref{fig:pedestrian-signals}a) vs. \textit{wait} (Figure~\ref{fig:pedestrian-signals}b). 
We experimentally fine-tuned the thresholds to identify the signal state. 

Our formative interviews revealed that BLV pedestrians struggle with pacing themselves while crossing streets (\textbf{C3}). To assist them, StreetNav informs users of the remaining crossing time. Its computer vision pipeline tracks signal cycle durations and maintains a timer that records signal state changes. By observing full cycles, StreetNav accurately monitors signal states and timings. Periodic timer updates ensure adaptability to changes in signal durations due to traffic management.

\begin{figure}[t]
    \centering
    \includegraphics[width=0.99\linewidth]{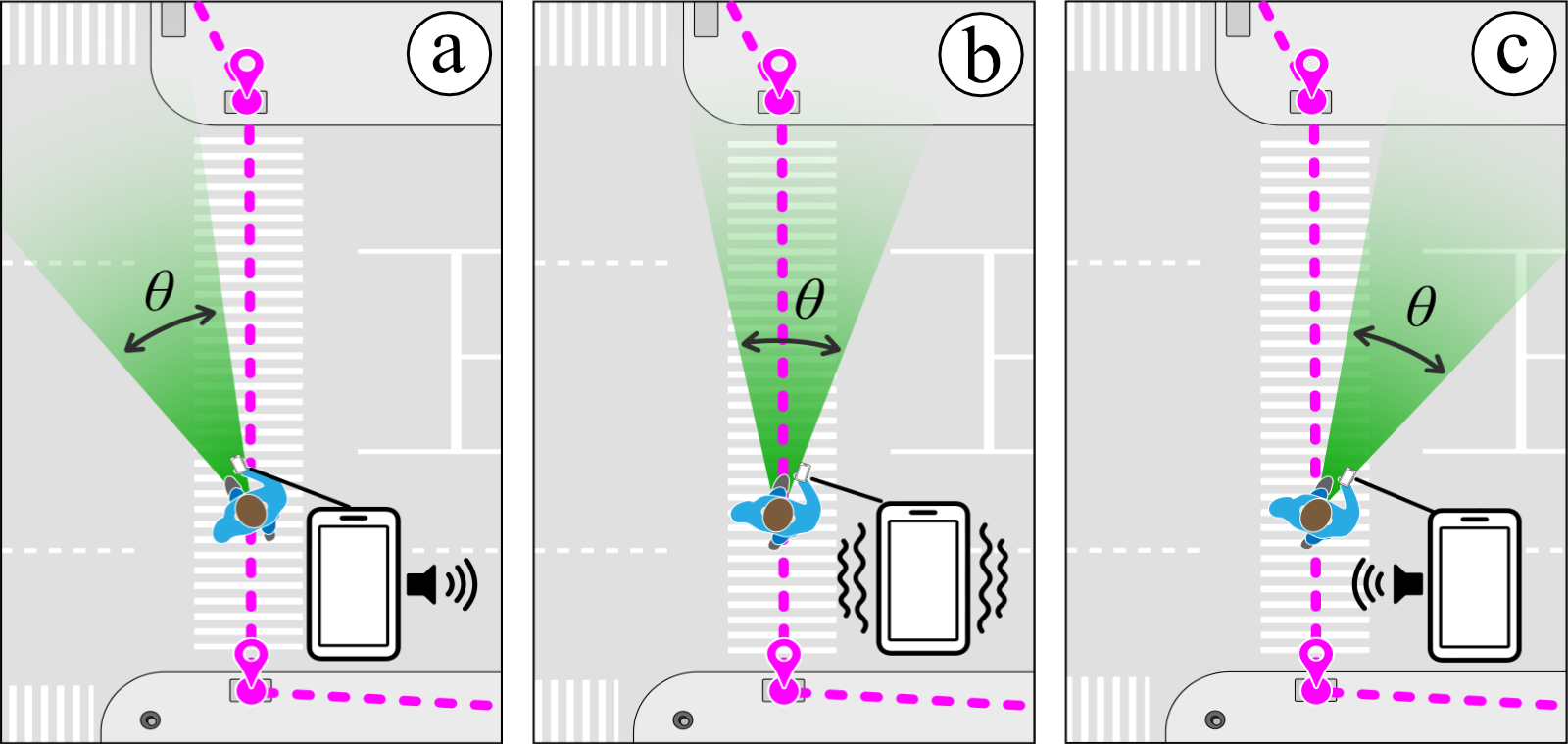}
         \caption{Audiohaptic cues for preventing users from veering off track. Sample user trajectories showing feedback when users (a) veers to the left, (b) do not veer, and (c) veer to the right. When the user's heading coincides with the route to the destination, within a tolerance angle $\theta$ (highlighted in green), users receive (b) subtle haptic vibrations to reinforce them. When they veer off the route, outside the tolerance angle $\theta$, they hear spatialized beeping sounds that are rendered from the (a) right speaker when veering left, and from the (c) left speaker when veering right.}
         \Description{Three part image labeled (a), (b), and (c), from left to right. Bird’s eye illustration of the crosswalk with the user crossing and their direction illustrated as a green vision cone coming from the front of the person, the calculated path shown as a pink dotted line straight across the street representing the ideal path, image of a phone in the lower right, illustrating the immediate haptic feedback. (a) Pedestrian on the pink path and green vision cone facing to the left of the path. An audio cue is coming from the right side of the phone. (b) Pedestrian on the pink path and green vision cone facing the direction of the path, haptic feedback coming from the phone. (c) Pedestrian on the pink path but green vision cone facing to the right of the path, audio cue coming from the left side of the phone.}
         \label{fig:veering}
\end{figure}

\begin{figure*}[t]
    \centering
    \includegraphics[width=0.951\linewidth]{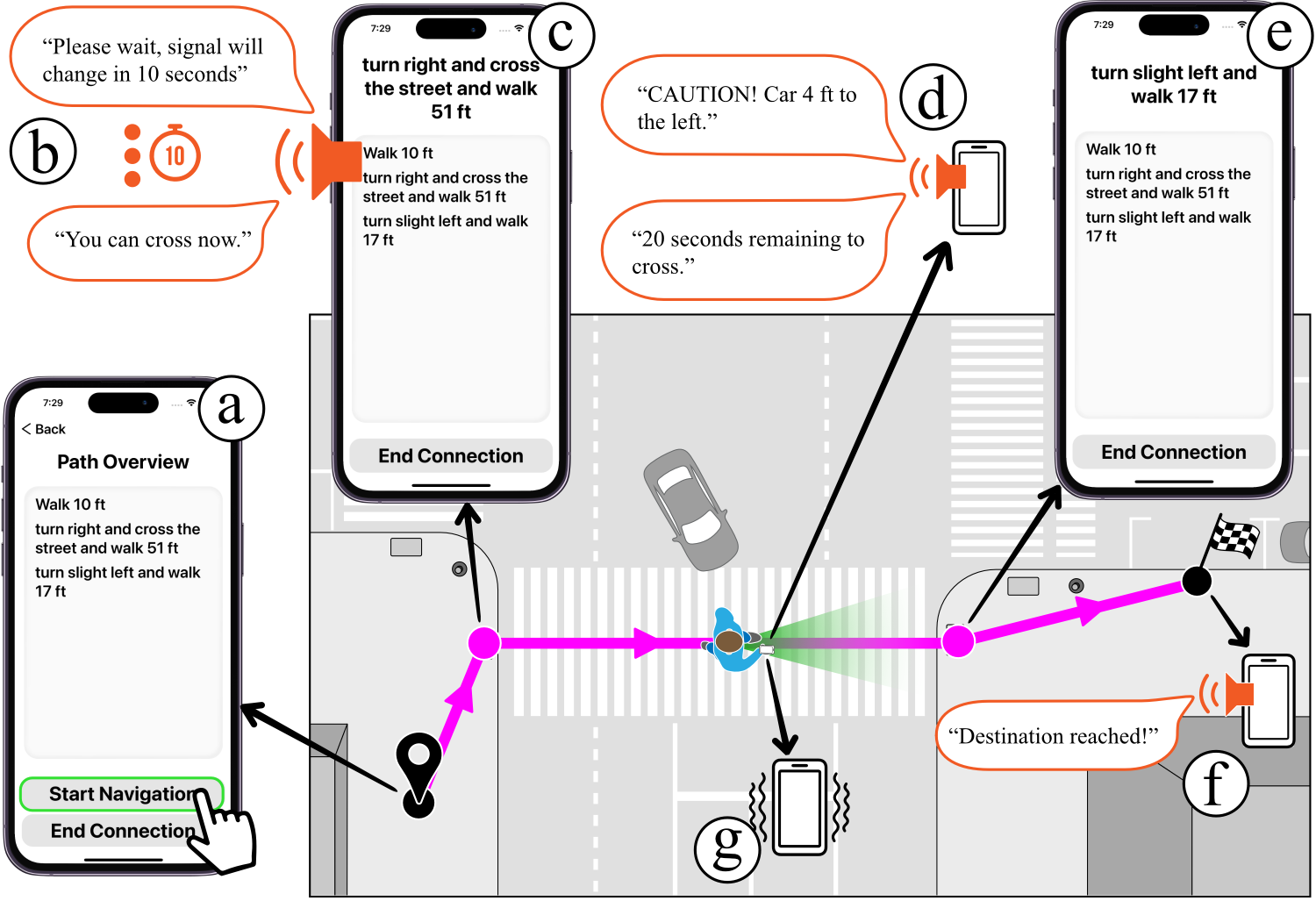}
         \caption{The StreetNav App's user interface. It provides routing instructions to their destination via (a) a path overview and (c, e) real-time feedback that updates their current instruction based on their location. Upon reaching a sidewalk, (b) the app informs the user about when it is safe to cross and (d) how much remains for them to cross over. It also (d) notifies the user of a nearby obstacle's category and relative location to help them avoid it. The app (f) confirms the user's arrival at the destination. Throughout the journey, the app provides (g) continuous audiohaptic feedback to prevent users from veering off track.
         }
         \Description{Simulated bird’s eye view of a street intersection with an overlaid pink path and waypoints that a blind pedestrian is traversing through. At every waypoint there’s a zoom in on a mobile app’s feedback. At the beginning of the path, the mobile phone shows a path overview. At the beginning of the cross walk, the mobile phone has an audio icon with a warning to the user that reads “Please wait, signal will change in 10 seconds”, and then “You can cross now”. In the middle of the crosswalk, the phone has an audio icon with a warning to the user that reads “Caution! Car 4 ft to the left!”. A car is seen yielding, in close proximity to the blind pedestrian. At the end of the path the mobile phone has an audio icon that reads “Destination reached!”. 
}
         \label{fig:app-ux}
\end{figure*}

\subsection{StreetNav App: User Interface}
\label{sec:user-interface}

The StreetNav iOS app interacts with the computer vision pipeline to allow BLV pedestrians to choose a destination and receive real-time navigation feedback that guides them to it. BLV users first initiate a connection request through the app, which activates the gesture-based localization (Section~\ref{sec:cv-pipeline}) in the computer vision pipeline. The app prompts the user to wave one hand over their head (Figure~\ref{fig:localization}b), enabling the system to begin tracking their precise location on the map (Figure~\ref{fig:localization}d). BLV users can then select a destination from nearby POIs and begin receiving navigation feedback through the app.

Figure~\ref{fig:app-ux} shows the StreetNav app's user interface, which uses audiohaptic cues for (i)~providing routing instructions, (ii)~preventing veering off track, (iii)~notifying about nearby obstacles, and (iv)~assisting with crossing streets. Upon reaching the destination, the app confirms their arrival. 
The following sections describe the app's interface in detail.


\subsubsection*{\textbf{Providing routing instructions}}
\rebuttal{The app provides routing instructions to users by offering a route overview before they start walking, as shown in Figure~\ref{fig:app-ux}a. This helps users prepare for their journey~\cite{guerreiro_virtual_2020, jain_i_2023, abd_hamid_facilitating_2013}. During navigation, the app announces instructions based on the user's location and provides continuous audiohaptic feedback to guide them.}

\rebuttal{Figure~\ref{fig:app-ux}b--f show how the app dynamically updates instructions based on the user's location. Users can access the path overview and current instructions on demand via VoiceOver~\cite{voiceover}.}

\rebuttal{Figure~\ref{fig:veering} illustrates the app's audiohaptic feedback. Based on the user's position, heading, and destination, StreetNav computes the direction and extent of veering. We initially used the Kalman filter to predict the user's heading based on their trajectory, but this proved inaccurate due to noisy tracking data. Instead, we used the smartphone's compass, offset by a fixed value to align its zero with the map's horizontal direction, allowing us to perform all heading computations relative to the map's frame of reference.}

\rebuttal{For directional guidance, we used stereo sound: beeping from the right speaker when users veer left (Figure~\ref{fig:veering}a) and from the left speaker when users veer right (Figure~\ref{fig:veering}c). The frequency of beeps increases with the extent of veering, allowing users to navigate effectively without headphones. To prevent overwhelming users with continuous audio feedback, a tolerance angle ($\theta$) of 50 degrees was introduced. Within this angle, subtle haptic vibrations guide users in the correct direction, while beeping sounds indicate veering, balancing audio as negative reinforcement and haptic feedback as positive reinforcement.}

\subsubsection*{\textbf{Notifying about nearby obstacles}}
Figure~\ref{fig:app-ux}d shows how StreetNav alerts the user of obstacles nearby. The app announces the obstacle's category, distance, and relative location. For example, when a car approaches the user, the app announces: ``\textit{Caution! Car, 4 ft. to the left.}'' Similar to veering feedback, the relative location is computed using both the computer vision pipeline's outputs and the smartphone's compass reading.

We tried feedback formats with varying granularity to convey the obstacle's relative location. First, we experimented with \textit{clock-faced directions}: ``\textit{Car, 4 ft. at 1 o'clock}.'' Clock-faced directions are commonly used in many GPS-based systems such as BlindSquare to convey directions. We learned from pilot evaluations with our BLV co-author that this feedback format was too fine-grained, as it took them a few seconds to decode the obstacle's location. This does not fare well with moving obstacles, such as pedestrians, that may have already passed the user before they are able to decode the location. Moreover, StreetNav's goal with obstacle awareness is to give users a quick idea that something is nearby them, which they can then use to circumnavigate via their mobility skills. To address this, we tried the more coarse format with just four directions: left, right, front, and back. This was found to give users a quick intimation, compared to the clock-faced directions.

\subsubsection*{\textbf{Assisting with crossing streets}}
The StreetNav app helps users cross streets by informing them \textit{when} to cross and how much time remains before the signal changes. 

Figure~\ref{fig:app-ux}b and Figure~\ref{fig:app-ux}d illustrate the feedback. Upon reaching a sidewalk corner, the app checks for the signal state recognized by the computer vision pipeline. If the signal is `\textit{wait}' when the user arrives, the app informs the user to wait along with the time remaining before the signal changes. If the signal is `\textit{walk}' when the user arrives, the app informs the user to begin crossing only if the time remaining is sufficient for crossing. For the intersection used in our user studies, this was experimentally found to be $15$ seconds. Otherwise, the user is advised to wait for the next cycle. Once the user begins crossing on the `\textit{walk}' signal, the app announces the time remaining for them to cross over. This feedback is repeated at fixed intervals until the user reaches the other sidewalk corner. We experimentally fine-tuned this interval with feedback from our BLV co-author. We tried several intervals, such as $5$, $10$, and $15$ seconds, and found that shorter intervals overwhelmed the users, whereas longer intervals practically would not be repeated enough times to give them meaningful information. We settled on repeating the feedback every $10$ seconds for our implementation.







\section{User Study}
\label{sec:user-eval}
Our user study had two goals, related to RQ3. 
First, we wanted to evaluate the extent to which StreetNav addressed BLV pedestrians' challenges in navigating outdoor environments when using existing GPS-based systems (Section~\ref{sec:formative-interviews}). 
Second, we wanted to analyze BLV pedestrians’ experience of navigating outdoors using StreetNav compared to existing GPS-based systems. 

\subsection{Study Description}
\subsubsection*{\textbf{Participants}}
We recruited eight BLV participants (five males, three females; aged 24--52) by posting to social media platforms and by snowball sampling~\cite{goodman_snowball_1961}. Participants identified themselves with a range of racial identities (Asian, Black, White, Latino, and Mixed), and all of them lived in a major city in the US. Participants also had diverse visual abilities, onset of vision impairment, and familiarity with assistive technology (AT) for navigation. 

Table~\ref{tab:main-study-ptcpts} summarizes participants' information. All but three participants (P1, P7, and P8) reported themselves as being moderately--extremely experienced with AT for navigation (3+ scores on a 5-point rating scale). Only P3 reported minor hearing loss in both ears and wore hearing aids. All participants except two (P2, P9) used white cane as their primary mobility aid. P2 did not use any mobility aid, while P9 primarily used a guide dog for navigation. The IRB-approved study lasted for about 120 minutes, and participants were compensated \$75 for their time. \revision{We obtained informed consent from all study participants.}

\begin{figure}[t]
    \centering
    \includegraphics[width=0.85\linewidth]{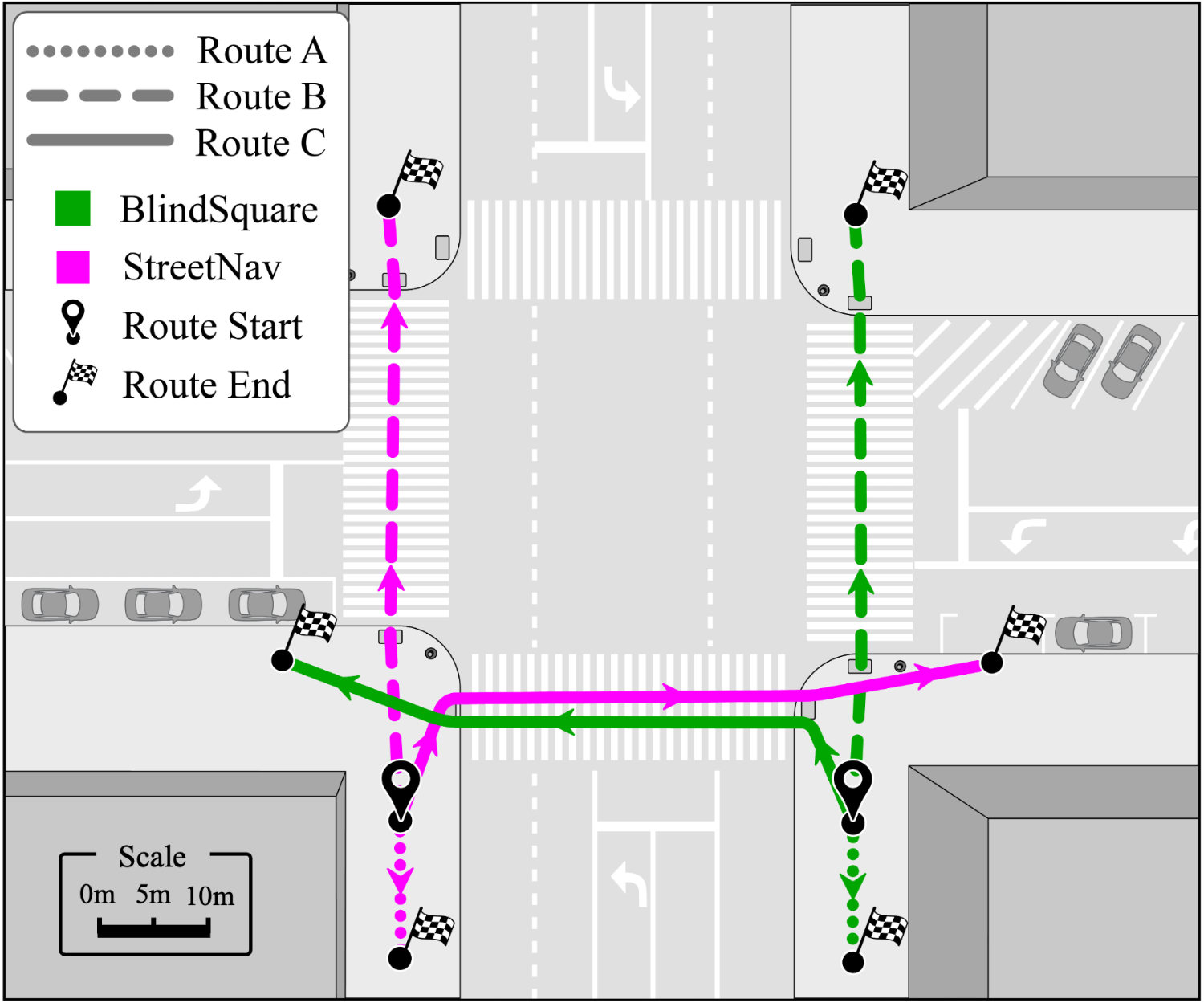}
         \caption{The routes used in the navigation tasks. (A) 12 meters, stationary person to avoid on the sidewalk. (B) 30 meters, cross street, and moving person to avoid on the sidewalk. (C) 38 meters, a $90^\circ$ turn, cross street, and moving person to avoid on the crosswalk. To mitigate learning effects, routes for the two conditions are symmetrically designed, situated on opposite sides of the street.
         }
         \Description{A bird’s eye illustration of the intersection with a legend in the upper left corner indicating the three types of paths and nodes on the illustration. There are three pink routes all starting from the same point, ending at different points, and three green routes all starting from the same point, ending at different points. The pink routes are StreetNav routes, the green routes are BlindSquare routes.}
         \label{fig:study-routes}
\end{figure}

\subsubsection*{\textbf{Experimental Design}}
In the study, participants completed three navigation tasks at a street intersection in two conditions: \textit{(i)} StreetNav and \textit{(ii)} BlindSquare~\cite{blindsquare}, a popular GPS-based navigation app specially designed for BLV people. We selected BlindSquare as the baseline because it emerged as one of the most frequently used apps among our BLV participants for outdoor navigation, as identified during the formative interviews (Section~\ref{sec:formative-interviews}). We evaluated the two systems via their respective iOS apps on an iPhone 14 Pro. Both systems' apps seamlessly integrated with VoiceOver, and all eight participants had a high level of familiarity with using iPhones and VoiceOver, with ratings of 3 or higher on a 5-point scale. 

\rebuttal{Note that our study objective is to compare StreetNav against BLV people’s current navigation methods using GPS-based systems. Since such apps, including BlindSquare, do not offer any assistance with obstacle awareness or crossing streets, the comparison effectively becomes StreetNav vs. participants' own abilities with mobility aids and non-visual senses.}

Our study followed a within-subjects design, in which participants tested the two navigation systems in a counter-balanced order to minimize potential order bias and learning effects. In each condition, participants were tasked with completing three distinct navigation challenges corresponding to three specific routes. Figure~\ref{fig:study-routes} illustrates these three navigation routes. We deliberately chose the routes to lie within the street camera's field of view and include a range of difficulty levels for each task: (A) a short route, 12 meters, that involved avoiding a stationary person on the sidewalk; (B) a long route, 30 meters, that involved crossing a street and avoiding a moving person on the sidewalk; and (C) a complex route, 38 meters, that involved making a $90$ degree turn, crossing a street, and avoiding a moving person on the crosswalk. For each of these tasks, one of the researchers assumed the role of an obstacle. None of the participants were familiar with the study location.

Given that participants navigated the same intersection in both conditions, the potential for learning effects as a confounding factor was carefully considered. To address this concern, we took deliberate measures by creating distinct routes for each condition. Specifically, we designed the routes in both conditions to be symmetric---rather than being identical---with the starting and ending points of each route strategically positioned on opposite sides of the street intersection, as illustrated in Figure~\ref{fig:study-routes}. The symmetry of routes ensured that participants encountered the same challenges in both conditions. To ensure participants' safety, the researchers accompanied them at all times during the study, prepared to intervene whenever necessary.

\subsubsection*{\textit{\textbf{Procedure}}}
We began each study condition by giving a short tutorial of the respective smartphone app for the system. During these tutorials, participants were taught how to use the app and how to interpret the various audiohaptic cues it offered. To accommodate potential challenges arising from ambient noise at the street intersection, participants were given the option to wear headphones during the study. Only two participants, namely P3 and P5, exercised that option; rest of the participants relied on the smartphone's built-in speaker to hear the audiohaptic cues.

After completing the three navigation tasks for each condition, we administered a questionnaire comprising four distinct parts. These parts were designed to assess participants' experiences around challenges faced by BLV pedestrians in outdoor navigation, specifically addressing the following aspects: routing to destination (\textbf{C1}), veering off course (\textbf{C1}), avoiding obstacles (\textbf{C2}), and crossing streets (\textbf{C3}). It included questions about how well each system assisted with the challenges, if at all. Participants rated their experience on a 5-point rating scale, where a rating of ``1'' indicated ``\textit{not at all well,}'' and a rating of ``5'' indicated ``\textit{extremely well.}'' After each part of the questionnaire, we asked follow-up questions to gain deeper insights into the reasons behind their ratings and their overall experiences.

Following their experience with both navigation systems, participants were asked to complete a post-study questionnaire. This questionnaire required them to rank the two navigation systems in terms of their preference for outdoor navigation. Subsequently, we directed our discussion toward StreetNav, engaging participants in a conversation about potential avenues for improvement. We also inquired about the specific scenarios in which they envision using this system in the future.

In addition to questionnaires capturing participants' subjective experiences, we also analyzed system usage logs and video recordings to assess participants' actual performance in the navigation tasks. We note that willingness to be video-recorded was completely voluntary. All eight participants still agreed to be video-recorded, providing us with written consent to do so.

\begin{figure}[t]
    \centering
    \includegraphics[width=0.99\linewidth]{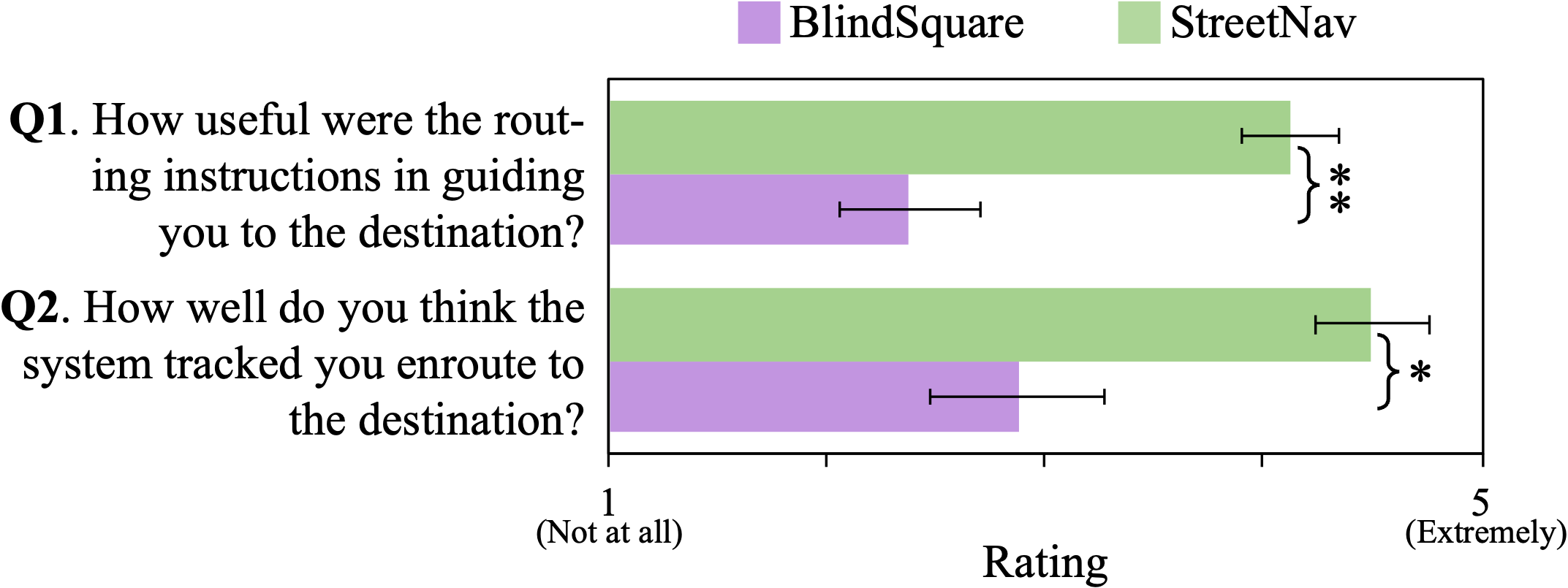}
    \caption{Results for participants' experience with routing to the destination. Participants rated the (1) usefulness of routing instructions, and (2) the system's ability to track them en route to the destination. Participants found StreetNav'sturn-by-turn instructions significantly more useful and precise than BlindSquare's ``as the crow flies''-style routing instructions. Pairwise significance is depicted for $p < 0.01$ ($*$) and $p < 0.05$ ($**$). The error bars indicate standard error.}
    \Description{Horizontal bar graph showing user reactions from 1 to 5 to the StreetNav system and to the BlindSquare system in user studies. Two questions, each question asked shows two bars, one green for StreetNav and one pink for BlindSquare. StreetNav  outscores Blindsquare on route instructions and enroute tracking to destination.}
    \label{fig:routing_results}
\end{figure}

\begin{figure*}[t]
    \centering
    \begin{subfigure}{0.4\linewidth}
        \centering
    \includegraphics[width=0.91\linewidth]{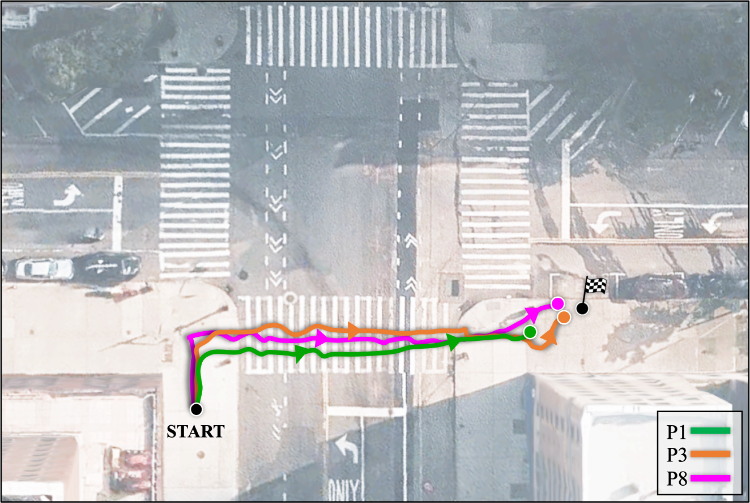}
        \caption{StreetNav} 
    \end{subfigure} 
    \hspace{-0.4cm}
    \begin{subfigure}{0.4\linewidth}
        \centering
        \includegraphics[width=0.91\linewidth]{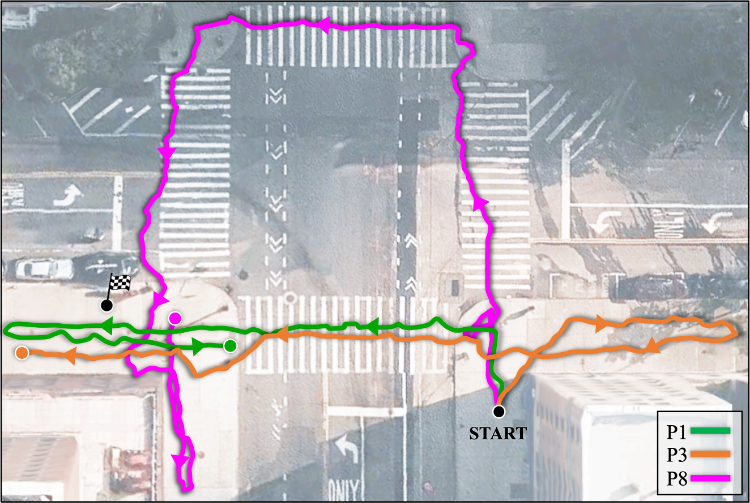}
        \caption{BlindSquare}
    \end{subfigure} 
    \caption{Comparison of paths traveled by three participants (P1, P3, P8) for route `C' using (a) StreetNav, and (b) BlindSquare. StreetNav's routing instructions consistently guided participants to the destination via the shortest path. BlindSquare, however, caused participants to take incorrect turns (P1, P3, P8), oscillate back and forth near destinations (P1, P8), and even go around the whole intersection before getting close to the destination (P8).}
    \Description{Two part bird’s eye view of the four way intersection with a legend in the lower right corner specifying three different path types; P1 green, P3 orange, P8 pink. Left is labeled (a) StreetNav and right is labeled (b) BlindSquare. (a) The three user paths follow the same route, very tightly grouped, across the crosswalk to the goal. (b) The three user paths follow wildly different routes around the intersection, and end up in three different locations but not the goal}
    \label{fig:route-comparison}
    
\end{figure*}


\subsubsection*{\textit{\textbf{Analysis}}}
We report participants’ spontaneous comments that best represent their overall opinions, providing further context on the quantitative data we collected during the study. We analyzed the transcripts for participants’ quotes and grouped them according to the (i) questionnaire's four parts: routing to destination, veering off course, avoiding obstacles, and crossing streets; (ii) overall satisfaction and ranking preferences, and (iii) how users' individual experiences influenced their preferences.

\subsection{Results} 
Our results show that StreetNav guided participants to destinations with greater precision and reduced veering, improved obstacle awareness, and increased confidence in street crossing. For the statistic analysis of each measure, we first used a Kolmogorov-Smirnov test to determine if the data was parametric. Then, for parametric data, we used a paired t-test to compare the two conditions. Additionally, we analyzed video recordings, annotating routes that participants took during the study. \rebuttal{We report key results below, with additional findings in Appendix~\ref{sec:addl-study-results}.}

\subsubsection*{\textbf{Routing to Destination}}

Figure~\ref{fig:routing_results} shows participants' average rating for their experience following routes to the destination in each condition. The mean ($\pm$ std. dev.) rating for participants' perceived usefulness of the routing instructions in guiding them to the destination was $4.13$ ($\pm 0.64$) for StreetNav and $2.38$ ($\pm 0.91$) for BlindSquare. The condition had a significant main effect ($p = 0.014$) on participants' experience reaching destinations with the routing instructions. The mean ($\pm$ std. dev.) rating for participants' experience with the system's ability to track them was $4.50$ ($\pm 0.76$) for StreetNav and $2.88$ ($\pm 1.13$) for BlindSquare. The condition had a significant main effect ($p = 0.001$) on participants' perception of how well the system tracked them en route to the destination. This indicates that participants found StreetNav more useful than BlindSquare for guiding them to the destination.

Figure~\ref{fig:route-comparison} illustrates our analysis of the video recordings, plotting the typical paths taken by participants in the third route across both conditions. We computed various metrics from their paths, that provide insights into participants' self-reported ratings. 

We found that when using BlindSquare, participants covered greater distances to reach the same destinations compared to when using StreetNav. On average, participants traveled a distance approximately $2.1$ times longer than the shortest route when relying on BlindSquare. In contrast, when using StreetNav, they covered a distance of only about $1.1$ times the shortest route to their destination. 
This represents a 51\% reduction in the unnecessary distance traveled with StreetNav in comparison to BlindSquare.
Figure~\ref{fig:route-comparison}b shows how participants using BlindSquare often exhibited an oscillatory pattern near their destinations (P1, P8) before eventually reaching close to them.

Additionally, StreetNav's routing instructions displayed a notably higher level of precision, guiding participants to their destinations with $2.9$ times greater accuracy than BlindSquare. Figure~\ref{fig:route-comparison} clearly shows this trend for the third route. On average, across the three study routes, participants using StreetNav concluded their journeys within a tighter radius of $12.53$ feet from their intended destination. In contrast, participants relying on BlindSquare concluded their journeys within a radius of $35.94$ feet from their intended destination. Two study participants, P4 and P5, even refused to navigate to the destination in two of the three tasks with BlindSquare. This was primarily attributed to BlindSquare's low precision in tracking the participants and often guiding them to take incorrect turns. Figure~\ref{fig:route-comparison}b highlights how BlindSquare caused P8 to go around the intersection before finally getting close the destination.

Participants preferred StreetNav over BlindSquare for its audiohaptic cues for turn-by-turn navigation instructions, which they found to be more useful and precise than BlindSquare's ``as the crow flies''-style clock face and distance-based instructions. P3's comment encapsulates this sentiment: \begin{quote} 
``\textit{When it's time for me to turn right and walk a certain distance, [StreetNav] is very, very, very precise.}'' --\textbf{P3}
\end{quote} 
Although all participants preferred StreetNav's routing feedback over BlindSquare's, distinct patterns emerged in their preference and utilization of these cues. 
StreetNav delivers a combination of audiohaptic and speech feedback for routing, and participants adopted varying strategies for utilizing this feedback.
Some individuals placed greater reliance on the veering haptic feedback as their primary directional guide, while reserving speech feedback as a fallback option. Conversely, some participants prioritized the speech feedback, assigning it a higher level of importance in their navigation process compared to audio-haptic cues.

\rebuttal{Maintaining a straight walking path is crucial for effective routing. Thus, we separately analyzed the extent to which each system prevented veering, with findings reported in Appendix~\ref{sec:veering-prevention-results}.}

\subsubsection*{\textbf{Obstacle Awareness}}
\begin{figure}[t]
    \centering
    \includegraphics[width=0.99\linewidth]{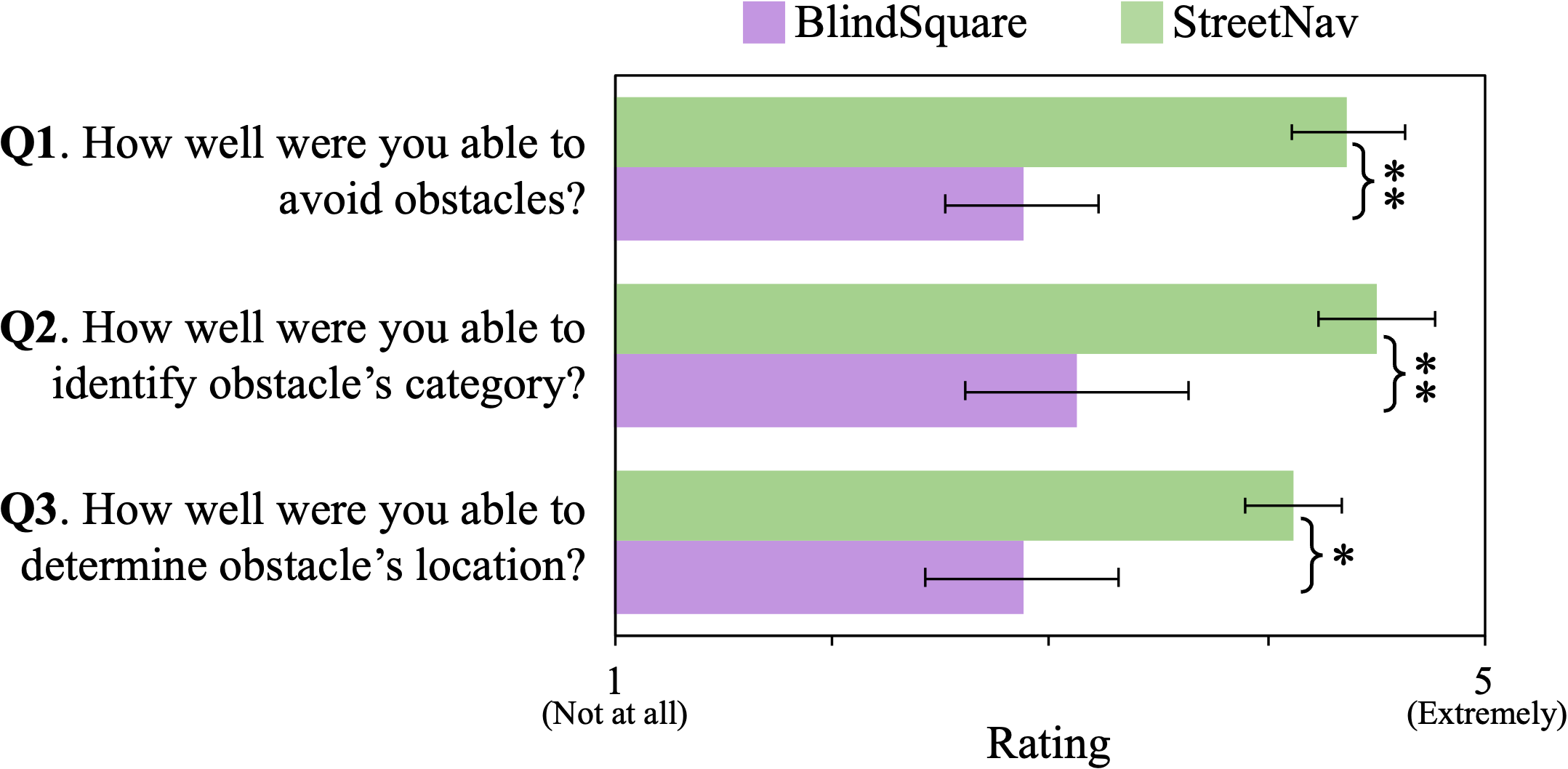}
    \caption{Results for participants' perceived obstacle awareness. Participants rated their ability to (1) avoid obstacles, (2) identify its category (e.g., person, bicycle), and (3) determine its relative location; on a scale of 1--5. StreetNav significantly improved participants' awareness of nearby obstacles during navigation. Pairwise significance is depicted for $p < 0.01$ ($*$) and $p < 0.05$ ($**$). The error bars indicate standard error.}
    \Description{Horizontal bar graph showing user reactions from 1 to 5 to the StreetNav system and to the BlindSquare system in user studies. Two questions, each question asked shows two bars, one green for StreetNav and one purple for BlindSquare. StreetNav scores higher on obstacle avoidance, identification, and localization tasks.}
    \label{fig:obstacle_results}
\end{figure}
Figure~\ref{fig:obstacle_results} shows participants' average rating for their perceived awareness of obstacles across the two conditions. Specifically, participants rated their ability to (1) avoid obstacles, (2) identify its category (e.g., person, bicycle, trash can), and (3) determine its relative location. The mean ($\pm$ std. dev.) rating for participants' perceived ability to avoid obstacles was $4.38$ ($\pm 0.74$) for StreetNav and $2.88$ ($\pm 0.99$) for BlindSquare, to identify its category was $4.50$ ($\pm 0.76$) for StreetNav and $3.13$ ($\pm 1.46$) for BlindSquare, and to determine obstacle's relative location was $4.13$ ($\pm 0.64$) for StreetNav and $2.88$ ($\pm 1.25$) for BlindSquare. A paired t-test revealed that the condition had a significant main effect on participants' perceived ability to avoid obstacles ($p = 0.030$), identify its category ($p = 0.037$), and relative location ($p = 0.004$). This suggests that StreetNav offered users a heightened awareness of nearby obstacles compared to the baseline condition of BlindSquare.

With StreetNav, participants had the option to use obstacle avoidance audio feedback in conjunction with their conventional mobility aids. However, in the case of BlindSquare, the system itself did not offer any obstacle-related information. Consequently, participants primarily relied on their traditional mobility aids in this condition, as is typical when using GPS-based systems. Our analysis of the video recordings found that in both experimental conditions, participants encountered no instances of being severely hindered by obstacles. Instead, they adeptly navigated around obstacles with the assistance of their white canes or guide dogs.

Although participants generally had a positive perception of obstacle avoidance when using StreetNav, their opinions on the utility of obstacle awareness information varied. Some participants found this information beneficial, emphasizing its role in preventing ``\textit{awkward bumping into people}'' (\textbf{P2}) and boosting their confidence, resulting in greater \textit{``speed in terms of walking''} (\textbf{P3}). Conversely, participants who felt confident avoiding obstacles with their mobility aids regarded StreetNav's obstacle information to be extraneous. P8 also expressed concerns about the potential information overload it could cause in dense urban areas: 
\begin{quote} 
\textit{``To know where people are, is a bit of overkill, because, especially in a city like this, if you turn this thing on in Times Square, it would have your head go upside down... If I'm around a lot of people, I'm not really thinking about avoiding them. I have a cane for a reason. They can see and I can't, so I'm relying on them to see me and get out of my way.''} --\textbf{P8}
\end{quote}

Many participants proposed an alternative use case for StreetNav's obstacle awareness information, highlighting its potential for providing insights into their surroundings. They suggested that this information could unlock environmental affordances, including the identification of accessible light signals and available benches for resting: ``\textit{knowing there was a bench was top-notch for me}'' (\textbf{P8}). Therefore, StreetNav's obstacle awareness information served a dual purpose, aiding in both obstacle avoidance and environmental awareness, allowing users to ``\textit{know what's around}'' (\textbf{P8}) them.

\subsubsection*{\textbf{Crossing Streets}}
\begin{figure}[t]
    \centering
    \includegraphics[width=0.99\linewidth]{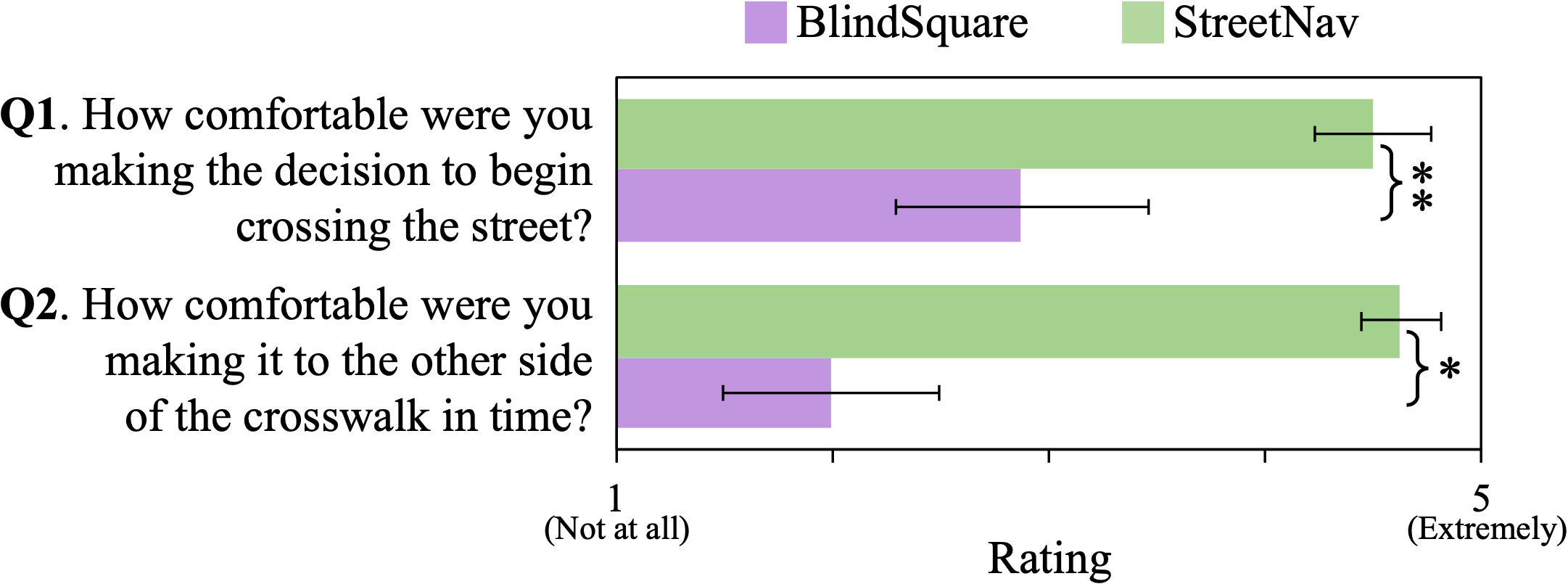}
    \caption{Results for participants' perceived comfort in crossing streets. Participants rated their perceived comfort in (1) making the decision on when to begin crossing the street, and in (2) pacing themselves when crossing. Participants were significantly more comfortable crossing streets with StreetNav in comparison to BlindSquare. Pairwise significance is depicted for $p < 0.01$ ($*$) and $p < 0.05$ ($**$). The error bars indicate standard error.}
    \Description{Horizontal bar graph showing user reactions from 1 to 5 to the StreetNav system and to the BlindSquare system in user studies. Two questions, each question asked shows two bars, one green for StreetNav and one pink for BlindSquare. StreetNav  outscores Blindsquare on crossing decision comfort and crossing time comfort.}
    \label{fig:cross_streets_results}
\end{figure}
Figure~\ref{fig:cross_streets_results} shows participants' average rating for their perceived comfort in crossing streets. The mean ($\pm$ std. dev.) rating of participants' perceived comfort in making the decision on when to begin crossing the street was $4.50$ ($\pm 0.76$) for StreetNav and $2.88$ ($\pm 1.64$) for BlindSquare. The mean ($\pm$ std. dev.) rating of participants' perceived comfort in safely making it through the crosswalk and reach the other end was  $4.63$ ($\pm 0.52$) for StreetNav and $2.00$ ($\pm 1.41$) for BlindSquare. A paired t-test showed that the condition had a significant main effect on participants' comfort in beginning to cross streets ($p = 0.029$) and in safely making it to the other side ($p = 0.001$).



As BlindSquare does not provide feedback for crossing streets, participants reported relying on their auditory senses by listening for the surge of parallel traffic. However, during the semi-structured interviews, some participants highlighted challenging scenarios that can make this strategy less reliable. P4, for instance, pointed out that ironically, less traffic can complicate street crossings: \begin{quote}
    \textit{``I don't always know when to cross because it's so quiet. And sometimes two, three light cycles go by, and I'm just standing there.''} --\textbf{P4}
\end{quote}
This issue has been exacerbated by the presence of electric cars, which are difficult to hear due to their quiet motors. For P3, their hearing impairments made it challenging to listen for traffic. Thus, most participants appreciated StreetNav's ability to assist with crossing streets: \begin{quote}
``\textit{When it's quiet, I would cross. But now with hybrid cars, it's not safe to do that. [StreetNav] app telling you which street light is coming on is really helpful.}'' --\textbf{P7}
\end{quote}

Participants made decisions to cross the streets by combining StreetNav's feedback with their auditory senses. Many participants emphasized that having information about the time remaining to cross significantly boosted their confidence, especially when this information aligned with the sounds of traffic: ``\textit{I thought it was great because I could tell that it matched up}'' (\textbf{P8}). This alignment between the provided information and their sensory perception inspired confidence in participants:
\begin{quote} 
``\textit{Relying on my senses alone feels like a gamble about 90 percent of the time, so a system like [StreetNav] that accurately displays the amount of time I have to cross the street is great.}'' --\textbf{P2}
\end{quote}
\rebuttal{P4 compared StreetNav with the Oko App~\cite{oko}. While P4 found Oko effective in identifying signal state, they appreciated StreetNav's seamless integration, which does not require pointing the camera at the pedestrian signal.}

\section{Technical Evaluation}
\label{sec:tech_eval}

\rebuttal{
We evaluate StreetNav's technical performance to compare its effectiveness against the status quo of GPS-based systems. StreetNav's main advantage is its precise user localization. Thus, this evaluation aims to answer the question: \textit{How precisely does StreetNav localize the user, and what factors impact this precision?}
}

\rebuttal{In comparing overall accuracy, StreetNav's localization error was $0.41$ ($\pm$ $1.49$) meters in estimating the user's feet position and an additional $0.65$ ($\pm$ $0.26$) meters in transforming this position from the camera view to the map. This error is significantly lower than GPS, which achieves localization errors in excess of $10$-$15$ meters in urban areas~\cite{gps_accuracy, modsching2006field, van_diggelen_worlds_2015}.
}

\rebuttal{We independently analyzed technical performance of the three steps in StreetNav’s computer vision pipeline for user localization: (i) CLIP-based gesture recognition, (ii) pedestrian feet position estimation, and (iii) camera to map-view transformation. Key results for each step are reported here, with detailed discussion in Appendix~\ref{sec:tech_eval_appendix}.}

\rebuttal{StreetNav's CLIP-based gesture recognition achieves 83\% accuracy in identifying the hand-waving gesture, with a 24\% false positive rate and a 10\% false negative rate. For pedestrian feet position estimation, the root mean squared error ($\pm$ std.) is $0.41$ ($\pm$ $1.49$) meters. Although StreetNav detects pedestrians with 82\% precision and 65\% recall at a 0.5 IOU (intersection over union) threshold, accuracy decreases as the pedestrian's distance from the camera increases, with the false negative rate rising from 1\% at $5$ meters to $74$\% at $40$ meters. The root mean squared error ($\pm$ std.) for transforming points from camera view to map view is $0.65$ ($\pm$ $0.26$) meters.}

\rebuttal{
Appendix~\ref{sec:tech_eval_appendix} elaborates on the evaluation procedure and provides additional detail on factors that impact performance for each step.}


\section{Discussion}






Our goal with StreetNav was to explore the idea of repurposing existing street cameras to support precise outdoor navigation for BLV pedestrians. We reflect upon our findings to discuss how street camera-based systems might be deployed at scale, privacy concerns with camera-based assistive technology, implications of a street camera-based navigation approach for existing GPS-based navigation systems, and the affordances enabled by precise, real-time outdoor navigation assistance.



\subsubsection*{\textbf{Deploying street camera-based navigation systems at scale}}

StreetNav demonstrates that street cameras have the potential to be repurposed for supporting precise outdoor navigation for BLV pedestrians. 
Our study results show that street camera-based navigation systems can guide users to their destination more precisely and prevent them from veering off course (Figure~\ref{fig:route-comparison}). 
Our results also show that street camera-based systems can support real-time, scene-aware assistance by notifying users of nearby obstacles (Figure~\ref{fig:obstacle_results}) and giving information about when to cross streets (Figure~\ref{fig:cross_streets_results}). These benefits of a street camera-based approach over existing GPS-based systems underscore the need for deploying such systems at scale. Although StreetNav was deployed at a single intersection, we learned insights on potential challenges and considerations that must be addressed to deploy street camera-based systems at scale.

Several internal and external factors need to be considered before street cameras can be effectively leveraged to support blind navigation at scale. 
External factors, including lighting conditions and occlusions on the street, may affect system performance. For instance, we noticed that StreetNav's ability to track pedestrians was affected severely in low-light conditions (e.g., at night) and by occlusions due to the presence of large vehicles (e.g., trucks, buses) and the installation of scaffolding for construction (Figure~\ref{fig:gesture-eval-failures}d). Such challenges affect the reliability of street camera-based systems and may limit its operational hours. Internal factors, including the positioning of cameras, their field of view, and variability in resolution, may affect the extent to which such systems can promise precise navigation assistance. For instance, the visibility of the pedestrian signals from the camera feed could affect how much such systems can assist users with crossing streets. With StreetNav, we observed a drop in tracking accuracy as pedestrians moved further away from the camera.


Therefore, deploying street camera-based systems at scale would require future work to investigate the extent to which both external factors (e.g., lighting, occlusions) and internal factors (e.g., camera resolution) affect system performance and reliability.
To address some of the technical limitations around tracking performance and field of view limitations, future research could explore integrating multiple cameras at various elevations and viewing angles. Prior work on robot navigation has explored the fusion of multiple cameras to improve tracking performance~\cite{chang_mobile_2013, oscadal_smart_2020, pflugfelder_localization_2010}.
Future work could also explore an ecosystem of accessible street cameras that can share information to automatically manage hand-offs across street intersections, providing users with a seamless experience beyond a single street intersection. Such ecosystems, which span beyond one intersection to a whole district or city, could enable new affordances, such as automatically sensing pedestrian traffic to inform traffic signals and vice versa~\cite{kostic_smart_2022}.

\subsubsection*{\textbf{Privacy concerns with camera-based assistive technology}}
\revision{
Privacy is a significant consideration for the practical deployment of street camera-based assistive technology. Our study with various stakeholders (Section~\ref{sec:formative-interviews-with-stakeholders}) revealed differing perspectives on privacy and identified strategies for respecting those perspectives. Recall from Section~\ref{sec:formative-interviews-with-stakeholders} the two strategies that our stakeholders identified: \textit{(i)} regulating data storage, anonymization, and access policies; and \textit{(ii)} repurposing \textit{existing} cameras rather than installing \textit{new} ones. Concerning the first strategy, StreetNav's implementation does not necessitate any data storage for facilitating outdoor navigation assistance. The video feed is processed in real-time on a local server, and only navigation instructions are shared with the BLV user's smartphone. Furthermore, StreetNav employs a map view representation---as depicted in Figure~\ref{fig:localization}d---for computing routes and identifying obstacles, inherently enabling data anonymization. The questions regarding who should have access to these cameras and for what other purposes, including public safety, they might be used for, still require further investigation. As for the second strategy, although StreetNav repurposes a camera from an existing publicly available testbed, the feasibility of securing camera access and resources of already existing street cameras at scale remains an open question. From our interview with the Community Board leader (Section~\ref{sec:formative-interviews-with-stakeholders}), collaboration among different government entities emerged as a potential next step. Future research could investigate the roles of different government entities and the implementation of policies that ensure responsible and transparent use of street cameras.
}

\subsubsection*{\textbf{Implications for GPS-based navigation systems}}
When cameras are available, and conditions align favorably, street camera-based systems offer BLV individuals a valuable source of fine-grained, high-precision information, significantly enhancing their navigational experience and environmental awareness. These capabilities are currently beyond the reach of conventional GPS-based systems. All eight study participants unanimously chose StreetNav over BlindSquare as their preferred navigation system due to its precise, scene-aware navigation assistance (Section~\ref{sec:forced-ranking}). 
However, it's important to acknowledge that street camera-based systems have their own set of limitations. The widespread availability of street cameras is not yet a reality, and ideal conditions may not always be met for their effective use. In contrast, GPS-based systems, while lacking in precision and environmental awareness, are universally accessible and resilient in varying conditions, including low light. 
A harmonious integration of these two approaches is a promising solution. Users can tap into street-camera information when conditions permit, seamlessly transitioning to GPS data when necessary. This can be facilitated through sensor fusion or information hand-offs, creating a synergy that ensures a smooth and reliable navigational experience. 
Future approaches could explore how these two systems can effectively complement each other, addressing their respective limitations and enhancing overall performance.

\subsubsection*{\textbf{Affordances of precise outdoor navigation assistance for BLV people}}
Previous research in indoor navigation has demonstrated the advantages of accurately pinpointing users' locations~\cite{sato_navcog3_2019, ahmetovic_navcog_2016, kim_navigating_2016} and providing scene-aware navigational information~\cite{guerreiro_cabot_2019, kayukawa_bbeep_2019}. However, achieving such precision has remained a challenge in outdoor environments, primarily due to the limited accuracy of GPS technology~\cite{gps_accuracy}.
StreetNav's approach of leveraging existing street cameras demonstrates that precise outdoor navigation support for BLV pedestrians is possible. Our study reveals the advantages of precise, fine-grained navigation for BLV individuals. These benefits include a substantial reduction in instances of veering and routing errors, such as deviation from the shortest path or missing intended destinations, as well as augmented environmental awareness. 

StreetNav offered our participants a glimpse into the potential of precise outdoor navigation. Several participants desired even greater precision, including the ability to discern the exact number of steps remaining before reaching a crosswalk's curb. Future research could delve into exploring how to best deliver such granular feedback to BLV users, alongside the necessary technological advancements needed to achieve this level of precision.
These advantages, as our findings suggest, extend beyond merely improving navigation performance. Participants shared insights into how precise navigation could enhance their independence when navigating outdoors. It could empower BLV people to venture outdoors more frequently, unlocking new travel opportunities, as exemplified by P3's newfound confidence in using public transportation with StreetNav-like systems: \begin{quote}
``\textit{I don't really use the city buses, except if I'm with somebody, but [StreetNav] would make me want to get up, go outside, and walk to the bus stop}.'' --\textbf{P3}
\end{quote}
This newfound confidence is particularly noteworthy, considering the unpredictable nature of outdoor environments. Future research could explore new affordances that street camera-based systems can enable for people, in general.

\section{Limitations}
Our work revealed valuable insights into the benefits and effectiveness of a new approach that uses existing street cameras for outdoor navigation assistance. At the same time, we acknowledge that our work has several limitations.

\revision{
StreetNav was developed using a camera from an existing cloud-networked testbed that is publicly available to the researchers~\cite{cosmos-cameras, yang2020cosmos,raychaudhuri2020challenge}, situated at a specific street intersection. It is important to note that our development process may not have encountered all potential technical challenges and design considerations, given the constraints of this setup. Additionally, StreetNav's use of the testbed camera instead of a regular security camera may yield slightly different performance due to factors like camera perspective, resolution, availability, and even the layout of the intersection.} Future research could expand upon our design and investigate how street camera-based systems can be adapted to different environments. 

Furthermore, to ensure the safety of participants and to fit the user study within a 120-minute timeframe, we designed the study routes to be less complex and dangerous. Real-world outdoor environments can vary significantly across regions, and our study location may not fully capture the diversity of scenarios BLV people encounter when navigating outdoors. 


Lastly, it is important to note that our design of StreetNav was guided by interviews with six BLV individuals, six stakeholders from New York City, and was evaluated in a study with only eight BLV individuals. While our participants' insights are valuable, their preferences may not represent the general population's perspectives on BLV people's navigation challenges and various stakeholders' privacy concerns. There could be additional challenges and design possibilities that we did not explore because of the cultural and regional context. Future research should consider a more extensive and diverse participant pool to gain a more comprehensive understanding of BLV people's challenges and privacy preferences of various stakeholders.

\section{Conclusion}
We explored the idea of leveraging \textit{existing} street cameras to support precise outdoor navigation for BLV pedestrians. Our resulting system, StreetNav, investigates both technical and sociotechnical concerns with a street camera-based navigation system. Our evaluations revealed StreetNav's potential to guide users more precisely to destinations compared to existing GPS-based systems. It also demonstrated camera-based system's ability to offer real-time, context-aware navigation assistance, aiding in obstacle avoidance and safe street crossings. However, we also identified challenges and opportunities for deploying street camera-based navigation systems at scale. These challenges suggest areas for future research to enhance system robustness and reliability while addressing privacy concerns. Our work highlights the potential of embedding accessibility into urban infrastructure using existing resources like street cameras. 
We envision a future where these systems seamlessly integrate into urban environments, providing BLV people with safe, precise navigation capabilities and empowering them to navigate their surroundings confidently.

\begin{acks}
We thank Lindsey Tara Weiskopf for literature review, Arjun Nichani for initial prototyping, and Chloe Tedjo and Josh Bassin for help with formative interviews. We thank our study participants for participating in the study. This work was supported in part by the National Science Foundation (NSF) and Center for Smart Streetscapes (CS3) under NSF Cooperative Agreement No. EEC-2133516, ARO Grant No. W911NF1910379, NSF Grant No. CNS-1827923, NSF Grant No. CNS-2038984, and corresponding support from the Federal Highway Administration (FHWA). Daniel Weiner and Xin Yi Therese Xu were supported by the Columbia--Amazon SURE Program. Sophie Ana Paris was funded by the NSF Grant No. 2051053 and 2051060.
\end{acks}

\balance
\bibliographystyle{ACM-Reference-Format}
\bibliography{main}
\section*{Appendix}
\appendix
\section{Formative Interviews with BLV People}
\label{sec:formative_study_appendix}
\rebuttal{
We provide details on the semi-structured interviews with BLV participants that we conducted to identify challenges that they face when navigating outdoors using GPS-based systems.
}

\begin{table*}[t]
\caption{Self-reported demographics of our participants. Gender information was collected as a free response; our participants identified themselves as female (F) or male (M). Participants rated their assistive technology (AT) familiarity on a scale of 1--5.
}
\label{tab:formative-study-ptcpts}
\renewcommand{\arraystretch}{1.25}
\begin{tabular}{cccclllll}
\hline 
\textbf{PID} & \textbf{Age} & \textbf{Gender} & \textbf{Race} & \textbf{Occupation} & \textbf{Vision ability} & \textbf{Onset} & \textbf{Mobility aid} & \textbf{AT familiarity (1--5)}
\\ \hline

F1  & 29 & Female & White & Claims expert & Totally blind & At birth &  White cane &   3: Moderately familiar \\ 

F2  & 61 & Female & White & Retired & Light perception only & Age 6 &  Guide dog &   1: Not at all familiar \\ 

F3  & 66 & Female & White & Retired & Totally blind & Age 58 &  Guide dog  &   2: Slightly familiar \\

F4  & 48 & Male & Black & Unemployed & Light perception only & Age 32 &  White cane &  3: Moderately familiar \\ 

F5 & 27 & Male & Mixed & Unemployed & Totally blind & At birth &  White cane &    3: Moderately familiar \\

F6 & 38 & Male & White & AT instructor & Totally blind & At birth &   White cane &    5: Extremely familiar \\

\hline
\end{tabular}
\end{table*}

\subsection{Methods}
\label{sec:f-methods}
\subsubsection*{\textbf{Participants}}
\rebuttal{
We recruited six BLV participants (three males, three females; aged 29--66) by posting on social media platforms and snowball sampling~\cite{goodman_snowball_1961}. Table~\ref{tab:formative-study-ptcpts} summarises the participants' information. All interviews were conducted over Zoom and lasted about 60 minutes. Participants were compensated \$25 for this IRB-approved study. We obtained informed consent from all study participants.
}

\subsubsection*{\textbf{Procedure}}
\rebuttal{
To identify the specific challenges that BLV people face when navigating outdoors, we used a recent critical incident technique (CIT)~\cite{flanagan_critical_1954}, in which we asked participants to recall and describe a recent time when they navigated outdoor environments using GPS-based assistive technology (AT). For example, we first asked participants to name the AT they commonly use and then asked them to elaborate on their recent experience of using it: \textit{``So, you mentioned using BlindSquare a lot. When was the last time you used it?''} Then, we initiated a discussion by establishing the scenario for them: \textit{``Now, let’s walk through your visit from the office to this restaurant. Suppose, I spotted you at your office. What would I observe? Let’s start with you getting out of your office building.''} We asked follow-up questions to gain insights into what made the aspects of outdoor navigation challenging and what additional information could help address them.
}

\subsubsection*{\textbf{Interview Analysis}}
\rebuttal{
To analyze the interviews, we first transcribed the study sessions in full and then performed thematic analysis~\cite{braun_using_2006} involving three members of our research team. Each researcher first independently went through the interview transcripts and used NVivo~\cite{nvivo} to create an initial set of codes. Then, all three iterated on the codes together to identify emerging themes.
}

\subsection{\textbf{Findings}}
\rebuttal{
We found three major themes around challenges that BLV pedestrians face when navigating outdoors using GPS-based systems.
}

\subsubsection*{\textbf{C1: Routing through complex environment layouts}}
\label{sec:challenge-1}
\rebuttal{
Participants reported difficulties in following routing instructions provided by GPS-based systems. These instructions, as explained by the participants, often did not match their current location. Many participants cited problems such as making wrong turns into unexpected ``alleyways'' (F1, F2, F4) that landed them in dangerous situations with ``cars coming through'' (F2). Participants cited examples of how these instructions caused them to veer off course---a common issue for BLV individuals in open, outdoor spaces~\cite{pan_walking_2013}---and end up in the middle of the streets.
This problem was particularly pronounced in complex environment layouts, as F3 recalled: \textit{``I didn't know if crosswalks were straight or curved or if they were angled. [It was hard] to figure out which way you needed to be to be in the crosswalk.''} Since \textit{"not everything is organized in the ideal grid-like way''} (F1), participants were hesitant to act on the navigation instructions without a clear understanding of the layout.
}

\subsubsection*{\textbf{C2: Avoiding unexpected obstacles while using GPS-based systems}}
\label{sec:challenge-2}
\rebuttal{
BLV people's challenges relating to obstacles during navigation are well researched~\cite{pariti_intelligent_2020, presti_watchout_2019}. However, we found specific nuances in their difficulties, particularly when they rely on their conventional mobility aids in conjunction with GPS-based navigation systems. Participants commonly reported the use of mobility aids like white canes alongside GPS systems for guidance. During this combined navigation process, they encountered difficulties in maintaining their focus on avoiding obstacles, often resulting in collisions with objects that they would have otherwise detected using their white canes. For instance, F2 shared an incident where they remarked, \textit{``there were traffic cones [and] I tripped over those''} while following directions from BlindSquare~\cite{blindsquare}. Notably, moving obstacles such as pedestrians and cars, as well as temporarily positioned stationary obstacles like triangle sandwich board signs, posed significant challenges for navigation. F4 expressed this sentiment, stating, \textit{``You know how many times I've walked into the sides of cars even though I have the right of way. Drivers have gotten angry, accusing me of scratching their vehicles. It can spoil your day [and make] you feel insecure.''}
}

\subsubsection*{\textbf{C3: Crossing street intersections safely}}
\label{sec:challenge-3}
\rebuttal{
Consistent with prior research \cite{guy_crossingguard_2012, mascetti_zebrarecognizer_2016, ahmetovic_mind_2017}, our study participants highlighted that crossing streets remained a significant challenge for them. Since GPS-based systems do not help with street-crossing, most participants relied on their auditory senses and apps like Oko~\cite{oko}. Regarding the use of auditory senses, they mentioned the practice of listening to vehicular sounds to gauge traffic flow on streets running parallel and perpendicular to their position. This auditory technique helped them assess when it was safe to cross streets. However, participants also reported instances where this method proved inadequate due to external factors: \textit{``yeah, it can be tricky, because [there may be] really loud construction nearby that can definitely throw me off because I'm trying to listen to the traffic''} (F1). Furthermore, their confidence in street-crossing decisions was affected by their inability to ascertain the duration of pedestrian signals and the length of the crosswalk. This uncertainty led to apprehension, as they expressed a fear of becoming stranded mid-crossing, as exemplified by one participant's comment: ``\textit{I don't want to be caught in the middle [of the street]}'' (F4). Regarding the use of Oko~\cite{oko}, participants found it cumbersome to point their phone's camera toward a pedestrian signal and to switch between this app and others during navigation.}

\section{Participant Demographics}
\label{sec:ptcpt-demographics}
\rebuttal{
Table~\ref{tab:stakeholder-study-ptcpts} summarizes demographics of various stakeholders we interviewed (Section~\ref{sec:formative-interviews-with-stakeholders}), and Table~\ref{tab:main-study-ptcpts} summarizes our user study participant demographics (Section~\ref{sec:user-eval}).
}

\begin{table*}[t]
\caption{Self-reported demographics of our user study participants. Gender information was collected as a free response. Participants rated their familiarity with assistive technology (AT) on a scale of 1--5.
}
\label{tab:main-study-ptcpts}
\renewcommand{\arraystretch}{1.3}
\begin{tabular}{cccclllll}
\hline 
\textbf{PID} & \textbf{Age} & \textbf{Gender} & \textbf{Occupation} & \textbf{Race} & \textbf{Vision ability} & \textbf{Onset} & \textbf{Mobility aid} & \textbf{AT familiarity (1--5)}
\\ \hline

P1 & 24	& Male	& App developer & Asian &	Low vision & Age 19	&  White cane & 2: Slightly familiar \\ 

P2 & 28	& Male	& Data manager	& White	& Low vision & 	At birth  & None  & 3: Moderately familiar \\

P3 & 48	& Male	& Not employed & Black & Totally blind & Age 32	& White cane & 3: Moderately familiar \\

P4 & 46	& Female	& Social worker	& Latino	& Totally blind	& Age 40	& White cane & 4: Very familiar \\

P5 & 43	& Female	& Not employed	& Asian	& Totally blind & 	At birth	& White cane &  4: Very familiar \\

P6 & 52	& Male	& Mgmt. analyst & Mixed	& Light perception only	& Age 9	& White cane & 5: Extremely familiar \\

P7 & 26	& Female & Writer & Mixed	& Low vision 	& At birth & White cane & 2: Slightly familiar \\


P8 & 51	& Male	& Not employed & Black	& Light perception only & Age 26 & Guide dog & 3: Moderately familiar \\

\hline
\end{tabular}
\end{table*}

\begin{table*}[t]
\caption{Self-reported demographics of our formative interviews with various stakeholders.}
\label{tab:stakeholder-study-ptcpts}
\renewcommand{\arraystretch}{1.32}
\begin{tabular}{clccl}
\hline 
\textbf{PID} & \textbf{Stakeholder Category} &  \textbf{Gender} &  \textbf{Age} & \textbf{Notes}
\\ \hline

B1 & BLV individual & Female & 62 & Light perception only \\

B2 & BLV individual & Gender Neutral & 41 & Limited vision in only left eye \\

R1 & Local resident & Female & 29 & Lived in Harlem for 12+ years \\

R2 & Local resident & Female & 35 & Lived in Harlem for 13+ years \\

O1 & Local business owner & Male & 58 & Running for 7+ years \\

CB1 & Community Board leader & Male & 53 & Serving as leader in Harlem \\

\hline
\end{tabular}
\end{table*}

\section{StreetNav: Technical Setup}
\label{sec:technical-setup}
\rebuttal{
Figure~\ref{fig:camera_location_view} shows the street camera we used for developing and evaluating StreetNav. 
The camera is part of the NSF PAWR COSMOS wireless edge-cloud testbed~\cite{raychaudhuri2020challenge, yang2020cosmos}, and is available to researchers after an approval process and IRB review.
We considered other publicly available testbeds such as Mobintel~\cite{mobintel} and DataCity SMTG~\cite{datacity_smtg}, but chose COSMOS due to its location in a major city (New York) with high pedestrian and vehicle traffic. Anonymized video samples from the COSMOS cameras, including the one used in this work, can be found online~\cite{cosmos-cameras}. StreetNav's computer vision pipeline takes the real-time video feed from the camera as input. 
For this purpose, we deployed the computer vision pipeline on one of the testbed servers, which captures the camera's video feed in real time. 
This server runs Ubuntu 20.04 with an Intel Xeon CPU@2.60GHz and an Nvidia V100 GPU. 
}

\rebuttal{
StreetNav's two components---the computer vision pipeline and the app---interact with each other via a cloud server, sharing information using the MQTT messaging protocol~\cite{mqtt}. Since MQTT is a lightweight messaging protocol, it runs efficiently even in low-bandwidth environments. The computer vision pipeline only sends processed navigation information (e.g., routing instructions, obstacle's category and location) to the app, rather than sending video data. This alleviates the privacy concerns around streaming the video feed to the users and avoids any computational bottlenecks that may happen due to smartphones' limited processing capabilities. The StreetNav app's primary purpose is to act as an interface between the user and the computer vision pipeline. We developed StreetNav's iOS App using Swift~\cite{swift}, enabling us to leverage VoiceOver~\cite{voiceover} and other built-in accessibility features.
}

\section{Additional User Study Results}
\label{sec:addl-study-results}

\subsection{Results for Veering Prevention}
\label{sec:veering-prevention-results}

\begin{figure}[t]
    \centering
    \includegraphics[width=0.99\linewidth]{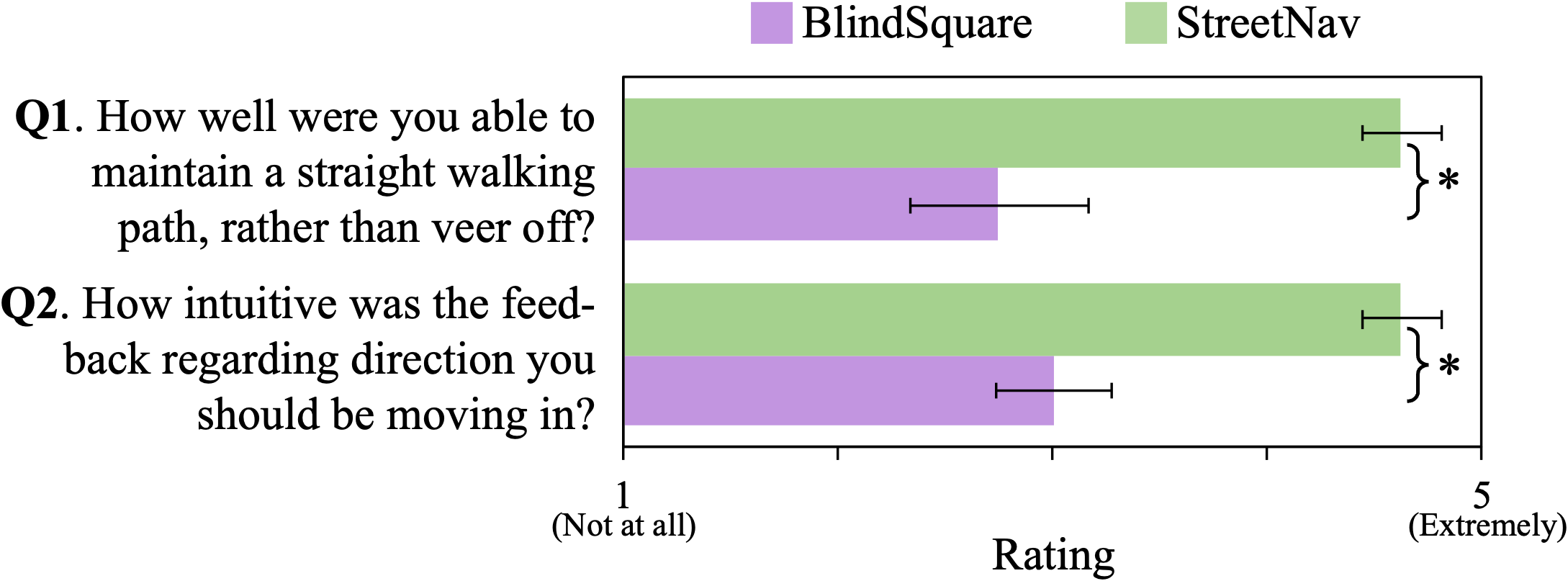}
    \caption{Results for participants' perceived ability to prevent veering off path. Participants rated their ability to (1) maintain a straight walking path, and (2) intuitiveness of the feedback regarding direction they should be moving in; on a scale of 1--5. StreetNav'saudiohaptic feedback was significantly more intuitive than BlindSquare's in preventing participants from veer off path. Pairwise significance is depicted for $p < 0.01$ ($*$). The error bars indicate standard error.}
    \Description{Horizontal bar graph showing user reactions from 1 to 5 to the StreetNav system and to the BlindSquare system in user studies. Two questions, each question asked shows two bars, one green for StreetNav and one pink for BlindSquare. StreetNav scores higher on maintaining straight walking path and veering feedback.}
    \label{fig:veering_results}
\end{figure}

Figure~\ref{fig:veering_results} shows participants' average rating for their perceived ability to (1) maintain a straight walking path, i.e., prevent veering off course, and (2) intuitiveness of the feedback they received regarding direction to move in. The mean ($\pm$ std. dev.) rating of participants' perceived ability to maintain a straight walking path with StreetNav was $4.63$ ($\pm 0.52$) and with BlindSquare was $2.75$ ($\pm 1.17$). The condition had a significant main effect ($p = 0.001$) on participants' perceived ability to prevent veering off course. The mean ($\pm$ std. dev.) rating for intuitiveness of the feedback that helped them know which direction to move in was $4.63$ ($\pm 0.52$) for StreetNav and $3.00$ ($\pm 0.76$) for BlindSquare. The condition had a significant main effect ($p = 0.006$) on intuitiveness of feedback that helped participants prevent veering off path.

Our examination of the video recordings aligns closely with participants' ratings. It reveals that StreetNav minimized participants' deviations from the shortest path to the destinations in comparison to BlindSquare. Over the course of the three routes, participants displayed an average deviation from shortest path, that was reduced by 53\% when using StreetNav as opposed to BlindSquare.


With BlindSquare, many participants reported difficulty maintaining awareness of their surroundings, including both obstacles and navigation direction, which frequently led to deviations from their intended paths. For instance, P2 reported challenges in maintaining their orientation with the need to avoid obstacles: \begin{quote} 
\textit{``[BlindSquare] basically demanded me to keep track of my orientation as I was moving, which is pretty difficult to do when you're also trying to keep other things in mind, like not bumping into things.''} --\textbf{P6}
\end{quote}

In contrast, StreetNav effectively addressed this challenge by providing continuous audiohaptic feedback for maintaining a straight walking path, instilling a sense of confidence in participants. P3, who tested StreetNav before BlindSquare, reflected on their desire for a similar continuous feedback mechanism within BlindSquare, akin to the experience they had with StreetNav: \begin{quote} 
\textit{``[with BlindSquare] even though I couldn't see the phone screen, my eyes actually went towards where I'm holding the screen. It is almost as if on a subconscious level, I was trying to get more feedback. With [StreetNav] I had enough feedback.''} --\textbf{P3}
\end{quote}
Many participants appreciated StreetNav's choice of haptic feedback for veering. Some participants envisioned the haptic feedback to be especially useful in environments with complex layouts:
\begin{quote}
``\textit{In the [areas] where the streets are very slanted and confusing. I think haptic feedback will be especially helpful.''} --\textbf{P5}
\end{quote}
Other participants highlighted the advantage of haptic feedback in noisy environments where audio and speech feedback might be less effective.

However, both P4 and P6 exclaimed that StreetNav's haptic feedback would only work well when holding the phone in their hands. This meant that hands-free operation of the app may not be possible, which is important for BLV people since one of their hands is always occupied by the white cane. P4 proposed integrating the app with their smartwatch for rendering the haptic feedback to enable hands-free operation.

\subsection{Forced Ranking Results} 
\label{sec:forced-ranking}

All eight participants unanimously chose StreetNav over BlindSquare as their preferred navigation assistance system. We asked participants to also rank their preferred type of routing instructions. All eight participants strongly preferred StreetNav's turn-by-turn routing instructions compared to BlindSquare's ``as the crow flies,'' direction and distance-style routing instructions.

In the semi-structured interview, participants were asked to elaborate on their rankings. Participants pointed out multiple navigation gaps in BlindSquare, with P2 summarizing participants' sentiment:
\begin{quote} 
``\textit{If you're only getting somebody 90 percent of the way there, you're not really achieving what I would consider to be the prime functionality of the system.}'' --\textbf{P2}
\end{quote}
In contrast, participants praised StreetNav for its precision and real-time feedback, emphasizing the importance of granular and holistic information to support all facets of navigation. However, participants did acknowledge occasional ``glitchiness'' (\textbf{P7}) with StreetNav, which occurred when they moved out of the camera's field of view or were occluded by other pedestrians or vehicles, resulting in lost tracking. Nevertheless, participants still regarded StreetNav as a significant enhancement to their typical navigation experiences, expressing increased confidence in exploring unfamiliar outdoor environments in the future.
\begin{quote}
``\textit{It would encourage me to do things that I would not usually... It would make me more confident about going out by myself.}'' --\textbf{P4}
\end{quote}
Participants also appreciated StreetNav's ability to identify them in near real-time: 
\begin{quote}
``\textit{What I found very interesting about the connection part is how quickly it identifies where I am, as soon as I waved my hand, it senses me.}'' --\textbf{P3}
\end{quote}

\rebuttal{
Participants also provided suggestions for improving StreetNav. Some participants wanted a hands-free version that would allow them to hold a white cane in one hand while keeping the other free. Additionally, while they found the gesture of waving hands for connecting with the system socially acceptable, they acknowledged that it might be perceived as somewhat awkward by others in the street. 
\begin{quote}
``\textit{[Waving a hand] may seem kind of weird to people who don't understand what is going on. But for me personally, I have no issue}.'' --\textbf{P3}
\end{quote}
Some participants highlighted that waving a hand might be misinterpreted by others on the street as a call for help, and may even cause security issues if a malicious person becomes aware that they were blind. P1 highlighted the role of public education in addressing this concern:
\begin{quote}
    ``\textit{If [others] see someone with a white cane, they know that's a blind person traveling. But if they see someone with their hand raised, they might think someone needs help or hailing a cab. So, I think it's going to be education to other people as much as to the person who is using this navigation system.}'' --\textbf{P1}
\end{quote}
}

\subsection{How Individual Experiences Influenced Participants' Preferences}
\rebuttal{
Throughout the study, participants offered feedback based on their unique backgrounds. We observed distinct patterns in their preferences, affected by their (i) onset of vision impairment, (ii) level of vision impairment, and (iii) familiarity with assistive technology.
}

\subsubsection*{\textbf{Onset of vision impairment}}
\rebuttal{Participants with early onset blindness preferred nuanced, concise feedback with an emphasis on environmental awareness. They used the system as an additional data point without complete reliance. In contrast, participants with late onset blindness trusted the system more and relied heavily on its feedback.}

\subsubsection*{\textbf{Level of vision impairment}}
Totally blind participants appreciated the veering feedback, while low-vision users, having more visual information, relied on their senses and needed less assistance with veering. Low-vision participants preferred the street-crossing feedback to interpreting pedestrian signals across the street. Totally blind participants primarily listened for parallel traffic, their usual method, using StreetNav's feedback for confirmation.

\subsubsection*{\textbf{Familiarity with assistive technology (AT)}}
\rebuttal{We noticed that participants who commonly use AT for navigation quickly adapted to StreetNav, while those with less experience hesitated in trusting StreetNav's feedback and had a slightly steeper learning curve. Still, all participants mentioned feeling more comfortable with StreetNav as the study progressed. Both groups also expressed increased confidence in exploring new areas with StreetNav.}

\begin{figure}[t]
    \centering
    \includegraphics[width=0.99\linewidth]{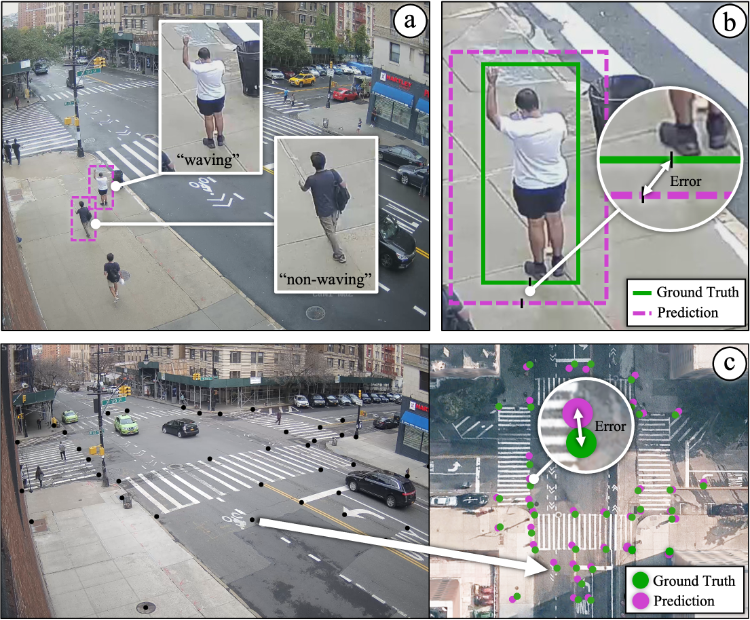}
         \caption{Illustration of StreetNav's localization steps analyzed in the technical evaluation: (a) CLIP-based gesture recognition, (b) pedestrian feet position estimation, (c) camera to map-view transformation.
         }
         \Description{Three part image labeled (a), (b), (c). (a) Top left is a blind man standing on the sidewalk, waving one hand above his head, with a phone in the other hand. (c) Bottom left is a screenshot of a phone screen prompting the user to wave. (b) Top right is a street intersection with bounding boxes around all of the pedestrians in the image. (d) Bottom right is the map view of the same intersection with circles indicating the positions of pedestrians.}
         \label{fig:tech-eval-flow}
\end{figure}

\section{Technical Evaluation}
\label{sec:tech_eval_appendix}
\rebuttal{We independently analyzed the technical performance of each of the three steps that enable StreetNav's computer vision pipeline to localize the user. Figure~\ref{fig:tech-eval-flow} illustrates the three steps: \textit{(i)} CLIP-based gesture recognition (Figure~\ref{fig:tech-eval-flow}a), \textit{(ii)} pedestrian feet position estimation (Figure~\ref{fig:tech-eval-flow}b), and \textit{(iii)} camera to map-view transformation (Figure~\ref{fig:tech-eval-flow}c). Recall from Section~\ref{sec:cv-pipeline}, StreetNav first distinguishes the BLV pedestrians from other pedestrians by recognizing the hand-waving gesture, then estimates their feet position as the mid-point of bounding box's bottom edge, and finally transforms their feet position from the camera view to the map.}


\subsection{Procedure}

We recorded a 15-minute evaluation video ($22500$ frames) from the camera feed to perform the technical evaluation. While recording this video, researchers posed as users navigating through the street intersection and played out different scenarios, such as waving hands and crossing streets. We also analyzed the errors for each of the three steps, revealing factors that impact StreetNav's ability to precisely determine a user’s position.

\subsection{Results}


\subsubsection*{\textbf{CLIP-based gesture recognition.}}
To evaluate the first step, we randomly sampled a balanced dataset of $140$ image crops from the evaluation video. Figure~\ref{fig:tech-eval-flow}a highlights the pedestrian image crops from each class. The CLIP-based gesture recognition module classifies each crop as waving or non-waving (i.e., walking, standing) pedestrian.

\begin{figure}[ht]
    \centering
    \includegraphics[width=0.45\linewidth]{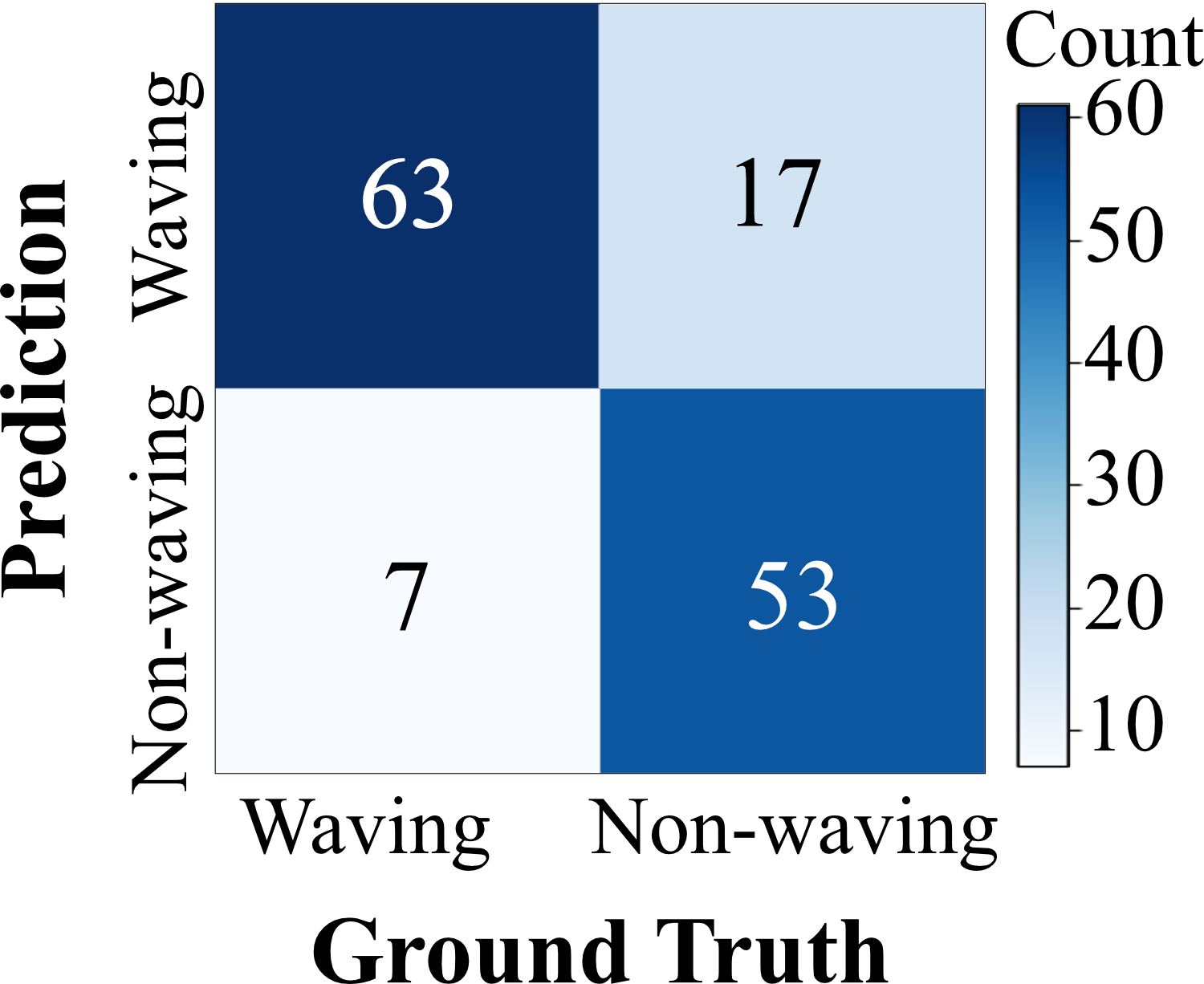}
         \caption{Confusion matrix for StreetNav's CLIP-based gesture recognition module. StreetNav distinguishes waving pedestrians from non-waving (i.e., walking, standing) ones with an 83\% accuracy.
         }
         \Description{}
         \label{fig:gesture-eval}
\end{figure}

\begin{figure}[ht]
    \centering
    \includegraphics[width=0.95\linewidth]{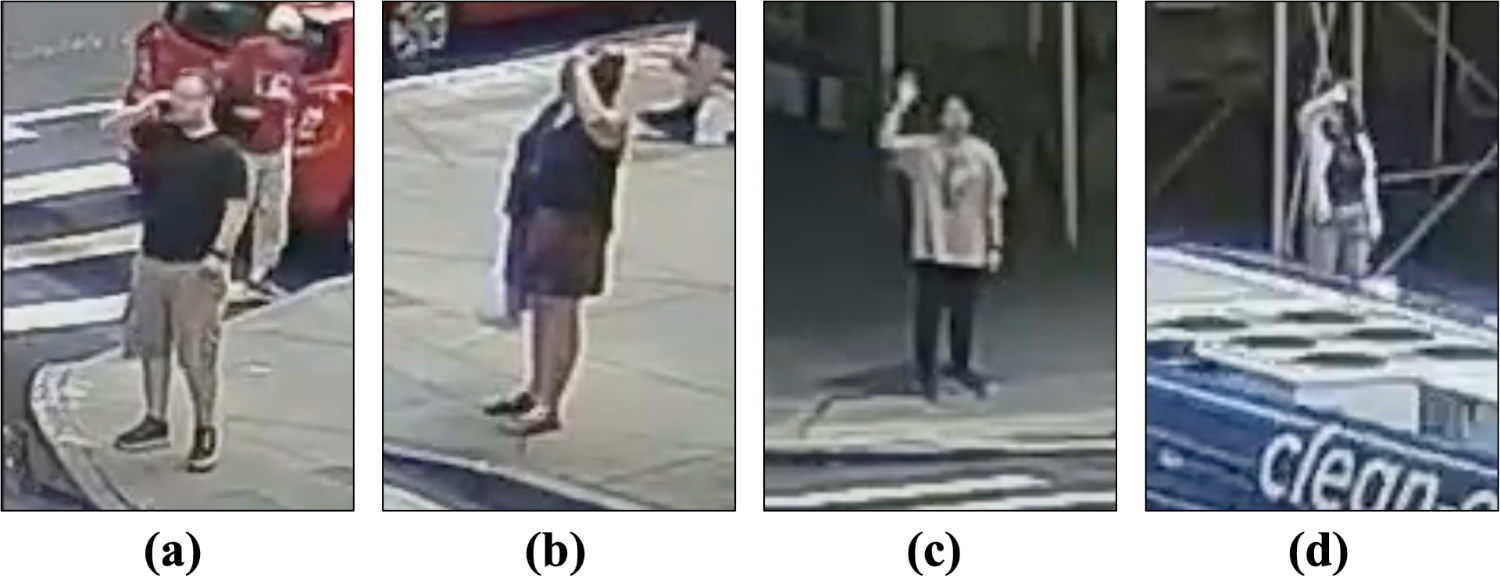}
         \caption{Failure cases in StreetNav's CLIP-based gesture recognition module. False positives occur when other pedestrians perform actions similar to waving their hand, such as (a) talking over phone or (b) casually resting their hand on forehead. False negatives occur when (c) users are too far from the camera and (d) due to foreground occlusions and background overlaps with vehicles, scaffolding, or other pedestrians.
         }
         \Description{}
         \label{fig:gesture-eval-failures} 
\end{figure}

Figure~\ref{fig:gesture-eval} shows the confusion matrix for CLIP-based gesture recognition module's performance. StreetNav achieves an $83\%$ accuracy in recognizing the hand-waving gesture, with a false positive rate of $24\%$ and a false negative rate of $10\%$. We analyzed the failure cases to identify specific scenarios that lead to the errors.

\rebuttal{Figure~\ref{fig:gesture-eval-failures} shows instances of the most common scenarios leading to false positives and false negatives. The false positives occur when other pedestrians perform actions similar to waving their hand, such as talking over a phone (Figure~\ref{fig:gesture-eval-failures}a) or casually resting their hand on their forehead (Figure~\ref{fig:gesture-eval-failures}b). The false negatives occur when users are too far from the camera (Figure~\ref{fig:gesture-eval-failures}c) or due to foreground occlusions and background overlaps such as vehicles, scaffolding, and other pedestrians (Figure~\ref{fig:gesture-eval-failures}d). While false negatives may result in users needing to wave their hands for a longer duration until recognized, false positives can lead them to follow incorrect instructions based on another pedestrian's location. StreetNav's approach to mitigating false positives is to announce the relative location of the detected pedestrian (e.g., `southwest corner'), providing users with additional contextual information to confirm whether they were recognized. The idea is that if this information does not align with the user's perception, they could then choose to re-establish the connection. Fine-tuning the CLIP model for this purpose could potentially enhance accuracy even further.
}

\subsubsection*{\textbf{Pedestrian feet position estimation.}}
\rebuttal{
To evaluate the second step, we manually annotated the ground truth pedestrian bounding boxes for $250$ frames, randomly sampled from the evaluation video. Figure~\ref{fig:tech-eval-flow}b shows the ground truth bounding box and StreetNav's predicted bounding box for a pedestrian. We report the root mean square errors between the feet positions estimated using the ground truth and predicted bounding boxes.}

\rebuttal{The root mean squared error ($\pm$ std.) in estimating pedestrians' feet position is $0.41$ ($\pm$ $1.49$) meters. The pixel distances were converted to physical distances to obtain the error in meters. We observed larger error rates for scenarios where pedestrians are occluded by other pedestrians or objects such as trash cans and fire hydrants. Future approaches could explore filtering abrupt changes in pedestrians' bounding boxes, caused by occlusions, to reduce this error.}

\rebuttal{
While analyzing the feet positions from the bounding boxes, we also noticed a trend in StreetNav's pedestrian detection pipeline. Recall from Section~\ref{sec:cv-pipeline}, StreetNav uses Nvidia's DCF-based multi-object tracker~\cite{nvidia_dcf_tracker} and the YOLOv8 object detector~\cite{terven2023comprehensive} for tracking pedestrians. We found that although StreetNav detects pedestrians with an 82\% precision and 65\% recall at 0.5 IOU (intersection over union) threshold, the accuracy drops significantly as the pedestrian's distance from the camera increases. This is attributed to the relatively smaller size of pedestrians, low resolution, and high chances of occlusion as pedestrians move further away from the camera. }

\begin{figure}[t]
    \centering
    \includegraphics[width=0.7\linewidth]{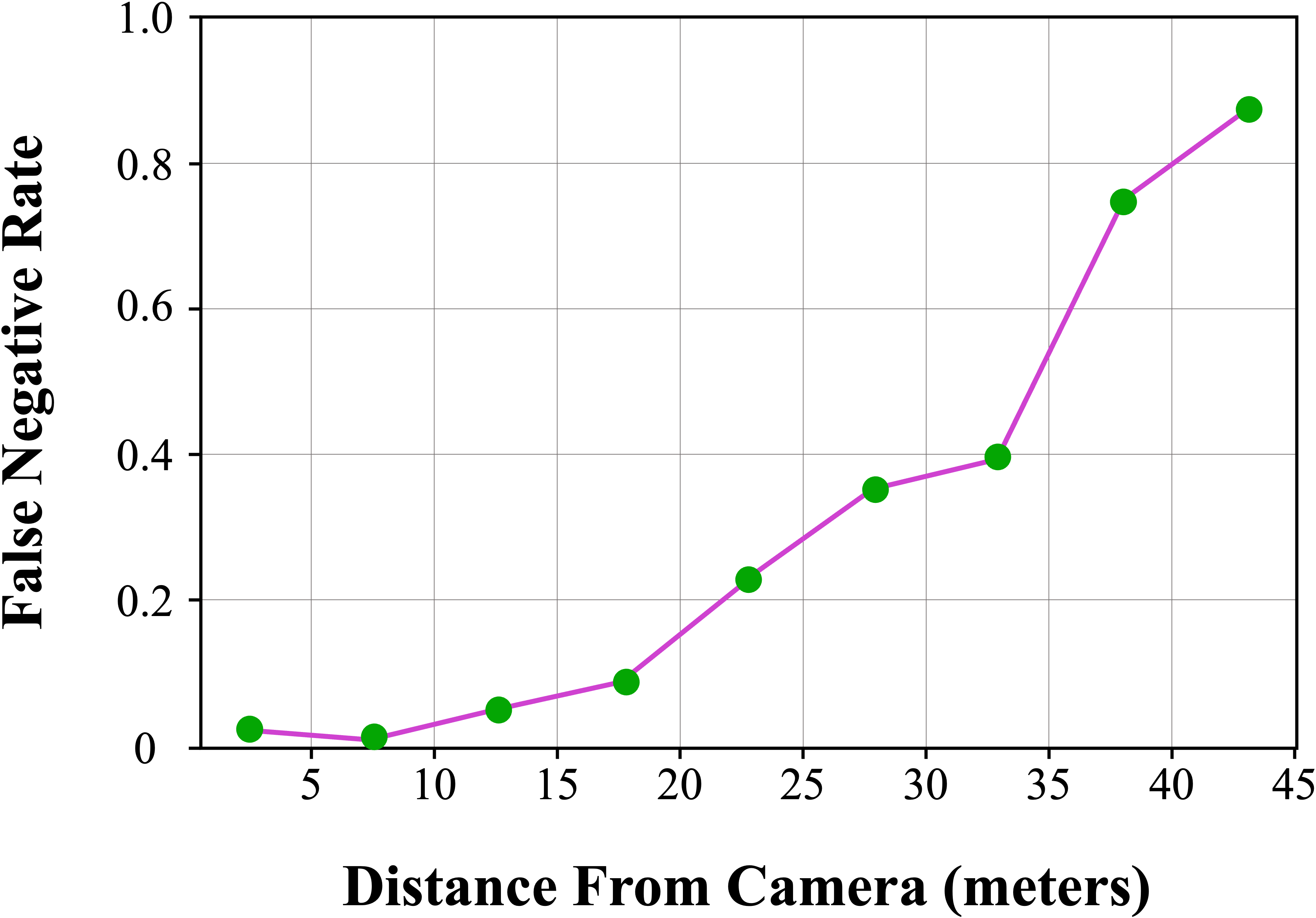}
         \caption{
         False negative rate (FNR) for pedestrian detection over distance from the camera in meters. StreetNav's error rates in detecting pedestrians increases significantly as they get further away from the camera. The FNR goes up from 1\% at 5 meters to 74\% at 40 meters distance from the camera.
         }
         \Description{}
         \label{fig:distance-camera-graph} 
\end{figure}

\rebuttal{
Figure~\ref{fig:distance-camera-graph} shows the false negative rate over distance from the camera. The false negative rate increases from 1\% at a distance of 5 meters from the camera to 74\% at a distance of 40 meters from the camera. Note that the distances were calculated between the pedestrian's feet estimations and the camera position's projection on the ground. Future approaches could combine detections from multiple cameras, such as two cameras positioned diagonally across a street intersection, to address this drop in accuracy. Alternatively, using training strategies that can detect both small and large pedestrians could also improve performance~\cite{hu_finding_2017}.
}

\subsubsection*{\textbf{Camera to map-view transformation.}}
\rebuttal{
To evaluate the third step, we selected a dataset of $50$ points in the camera view and we manually annotated their corresponding position on the map. We chose these specific points for evaluation as they correspond to visual landmarks on the street and are evenly spread across the street intersection. For example, we selected points on the crosswalk edges and road signs. As a result, annotating their ground truth position on the map view could be done with reasonable accuracy by simply comparing the camera and map view images. For these $50$ points, we also generated StreetNav's predicted transformations from the camera view to the map view. Figure~\ref{fig:tech-eval-flow}c shows the points we selected and their corresponding ground truth and predicted transformations. We computed the root mean square errors between the transformed ground truth positions and StreetNav's predicted positions.}

\rebuttal{The root mean squared error ($\pm$ std.) in transforming points from the camera view to the map view, averaged across the points shown in Figure~\ref{fig:tech-eval-flow}c, is $0.65$ ($\pm$ $0.26$) meters. The pixel distances were converted to physical distances to obtain the error in meters. These errors occur due to the curvature in the camera lens.
}



\end{document}